\documentclass[]{aa}

\usepackage{txfonts}
\usepackage{float}
\usepackage{graphicx}
\usepackage{subfigure}
\usepackage{tabularx}
\usepackage{epsfig}
\usepackage[usenames,dvipsnames]{xcolor}

\begin{document}

\title{Two spectroscopically confirmed galaxy structures at z=0.61 and 0.74 in the
  CFHTLS Deep~3 field }

\author{
C.~Adami\inst{1} \and
E.S.~Cypriano\inst{2} \and
F.~Durret\inst{3} \and
V.~Le~Brun\inst{1} \and
G.B.~Lima Neto\inst{2} \and
N.~Martinet\inst{3} \and
F.~Perez\inst{1} \and
B.~Rouze\inst{1} \and
L.~Sodr\'e~Jr.\inst{2}
}

\offprints{Christophe Adami \email{christophe.adami@lam.fr}}

\institute{
  Aix Marseille Universit\'e, CNRS, LAM (Laboratoire d'Astrophysique
  de Marseille) UMR 7326, 13388, Marseille, France
\and
IAG, USP, R. do Mat\~ao 1226, 05508-090, S\~ao Paulo, SP, Brazil 
\and
UPMC-CNRS, UMR7095, Institut d'Astrophysique de Paris, F-75014, Paris, France 
}

\date{Accepted . Received ; Draft printed: \today}

\authorrunning{Adami et al.}

\titlerunning{Two spectroscopically confirmed clusters in the CFHTLS D3 field}

\abstract 
{Galaxy evolution is known to depend on environment since it differs
  in clusters and in the field, but studies are sometimes limited to the
  relatively nearby Universe ($z<0.5$). It is still necessary to
increase our knowledge of cluster galaxy properties above $z=0.5$.}
{In a previous paper we have detected several cluster candidates at $z>0.5$ as part of
  a systematic search for clusters in the Canada France Hawaii
  Telescope Legacy Survey by applying the Adami \& MAzure Cluster
  FInder (AMACFI), based on photometric redshifts. We focus here on
  two of them, located in the Deep~3 (hereafter D3) field: D3-6 and D3-43.}
{We have obtained spectroscopy with Gemini/GMOS instrument and measured redshifts
  for 23 and 14 galaxies in the two structures. These redshifts were
  combined with those available in the literature. A dynamical and a weak 
  lensing analysis were also performed, together with
  the study of X-ray Chandra archive data.}
{Cluster D3-6 is found to be a single structure of height spectroscopically
  confirmed members at an average redshift $z=0.607$, with a velocity
  dispersion of 423~km~s$^{-1}$. It appears to be a relatively
    low-mass cluster.
  D3-43-S3 has 46 spectroscopically confirmed members at an average
  redshift $z=0.739$.  The cluster can be decomposed into two main
  substructures, having a velocity dispersion of about 600 and
  350~km~s$^{-1}$. An explanation of the fact that D3-43-S3 is
  detected through weak lensing (only marginally, at the
    $\sim$3$\sigma$ level) but not in X-rays could be that the two
  substructures are just beginning to merge more or less
  along the line of sight.
  We also show that D3-6 and D3-43-S3 have similar global galaxy
    luminosity functions, stellar mass functions, and star formation
    rate (SFR) distributions. The only differences are that D3-6
    exhibits a lack of faint early-type galaxies, a deficit of
    extremely high stellar mass galaxies compared to D3-43-S3, and an
    excess of very high star formation rate galaxies.}
{This study shows the power of techniques based on photometric
  redshifts to detect low to moderately massive structures, even at z$\sim$0.75. 
  Combined-approach cluster surveys such as EUCLID are
  crucial to find and study these clusters at these relatively high
  redshifts.  Finally, we show that photometric redshift
  techniques are also well suited to study the galaxy content and
  properties of the clusters (galaxy types, star formation rates,
  etc...).}


\maketitle

\section{Introduction}

Galaxy evolution is still a major topic in cosmology, because of the
complexity of the processes involved. The current theoretical
framework (e.g. White \& Frenk 1991) assumes that galaxies are formed
by accretion of baryonic matter onto dark matter halos and evolve
through mergers and other interactions with other galaxies and the
intergalactic medium (IGM). The study of evolution in clusters is
particularly interesting because some extreme conditions (e.g. high
galaxy density, large amounts of hot gas in the IGM) 
are found in these structures
and, consequently, galaxy evolution is affected by many
environmentally-driven processes, as evidenced by the evolution of the
morphological mix of their galaxies (e.g. Boselli et al. 2014). 
Naturally, in this scenario the
evolution of galaxies is not independent from the evolution of cluster
structures. According e.g. to Mateus et al. (2007), galaxy evolution is
accelerated in denser environments. All this makes clusters ideal
laboratories to investigate galaxy evolution.

Most of our current knowledge about processes affecting galaxies in
clusters comes primarily from studies in the nearby ($z < 0.5$)
Universe. At higher redshifts, several studies have focussed on the
evolution of morphological type fractions (e.g. Holden et al. 2007,
van der Wel et al. 2007, Simard et al. 2009), on their clustering
properties (e.g. Poggianti et al.  2010, Ross et al. 2010), or on the
evolution of specific galaxy types (such as red galaxies:
S\'anchez-Bl\'azquez et al. 2009).

The question we want to answer is how do cluster galaxies evolve in
the redshift interval $0.5 < z < 1$.  We propose adding our contribution to
this problem through spectroscopy and photometry of candidate cluster galaxies
observed with GEMINI/GMOS and primarily detected from CFHTLS survey
photometric redshift catalogues. In this framework, Adami et
al. (2010) have published a catalogue of $\sim$1200 cluster candidates from
public photometric redshifts (obtained following Ilbert et al. 2006;
 also see Coupon et al. 2009) of the CFHTLS data release T0004 in the
D2, D3, D4, W1, W3, and W4 regions. This catalogue contains several
cluster candidates at redshift larger than 0.5, detected with high confidence
(S/N equal to or larger than 4) and also detected by other methods
(Olsen et al. 2007, Thanjavur et al. 2009). We choose to focus on two
of them.

We describe the optical identification of our candidate clusters in
Section 2.  The mass characterisation of the two structures is
discussed in Section 3. In Section 4 we present some properties of the
galaxy populations in the clusters, and in Section 5 we summarise our
results.

We adopted, where necessary, 
the Dunkley et al. (2009) $\Lambda$CDM concordance cosmological model
(H$_0$=71.9~km~s$^{-1}$~Mpc$^{-1}$, $\Omega _\Lambda = 0.742$,
$\Omega_M=0.258$).

\section{Optical identification of our candidate clusters}

\subsection{CFHTLS imaging}

The two clusters studied here were discovered as part of a systematic
search for clusters in the Canada France Hawaii Telescope Legacy
Survey (Adami et al. 2010). They were found in the Deep~3 field, which
covers 1~deg$^2$ and is centred at coordinates 14:19:27.00, +52:40:56
(J2000.0). This field is among the deepest optical images available from 
the ground, corresponding to a total exposure time of the order of 90 hours.

To identify clusters in this field we applied
the  Adami \& MAzure cluster FInder (AMACFI, Mazure et
al. 2007), based on photometric redshifts (hereafter photo$-z$s).
After the photo$-z$ catalogue is cut in photometric redshift slices,
galaxy density maps are drawn with an adaptive kernel technique, and
overdensities are detected at a chosen significance level with
SExtractor. Detections are then sorted with a minimal spanning tree
method, leading to a catalogue of candidate clusters.

The two candidate clusters studied here were detected at a 4$\sigma$ significance level. 
Their identifications in the Adami et al. (2010) catalogue are D3-6 and D3-43.

\subsection{Gemini GMOS spectroscopy}

We were awarded 8.5 hours of GEMINI/GMOS time (program GN-2011A-Q46,
PI: L. Sodr\'e) to observe these two cluster
candidates spectroscopically. We initially used the R400 grism and one arcsec slits. 
Final 1D spectra were rebinned to have $\sim$7 \AA~ per pixel.

We were able to fit 27 and 16 slits in the D3-6 and D3-43
fields respectively. Reduction was made in the IRAF environment with the GMOS
dedicated tools.  We applied the EZ redshift measurement code (Garilli
et al. 2010) on the final 1D spectra, allowing an additional smoothing
from 3 to 9 pixels to find the redshift
value more easily. The redshift measurements were done in the same way as for the
VIPERS survey (e.g. Guzzo et al. 2014). Independent measurers provided
two first estimates of the redshifts (V. Le~Brun and C. Adami). The
two values were then reconciled and a quality flag was assigned
between 1~and 4. Flag 1 means that we have a 50$\%$ chance to have
the correct redshift estimate, flag 2 means that we have a 75$\%$ chance, 
flag 3 means that we have a 95$\%$ chance, and flag
4 means that we have more than 99$\%$ chance.  We only considered the
objects with flags 2, 3, and 4 to be successful measurements.  We obtained an excellent success
rate: 15 of the 16 D3-43 spectra and 26 of the 27 D3-6 spectra
provided successful redshift measurements. Only four spectra turned
out to be of stellar origin, while we obtained galaxy redshifts for 23
and 14 galaxies in the fields of D3-6 and D3-43, respectively.

Coordinates and successfully measured redshifts for this sample are
given in Tables \ref{tab:sample1} and \ref{tab:sample2}. We show four
examples of spectra corresponding to flags 4, 3, and 2 in
Fig.~\ref{fig:example1} and ~\ref{fig:example2}.

\begin{table}[t!]
  \caption{Running number, coordinates (J2000), redshifts, and spectral flags for the objects observed 
    with GEMINI/GMOS in the D3-43 field of view.}
\begin{center}
\begin{tabular}{rrrrr}
\hline
\hline
\#  & RA           &  DEC        & redshift & flag \\
\hline
1 & 14:20:51.079 & 53:03:38.76 & 0.8264 & 3 \\
2 & 14:20:56.214 & 53:02:21.28 &  0.8950 & 3 \\
3 & 14:20:56.472 & 53:02:35.13 &  0.7111 & 4 \\
4 & 14:20:58.824 & 53:01:11.63 &  0.7497 & 4 \\
5 & 14:20:59.161 & 53:01:36.41 &  0.7404 & 4 \\
6 & 14:21:00.276 & 53:04:29.72 & 0.7737 & 4 \\
7 & 14:21:02.344 & 53:05:03.26 & 0.6723 & 4 \\
8 & 14:21:02.572 & 53:01:22.74 &  0.3347 & 4 \\
9 & 14:21:05.073 & 53:02:06.92 & 0.7393 & 4 \\
10& 14:21:10.762 & 53:02:49.93 & 0.7350 & 4 \\
11& 14:21:11.665 & 53:04:44.20 & 0.8217 & 3 \\
12& 14:21:15.013 & 53:00:57.71 &  0. & 4 \\
13& 14:21:18.394 & 53:03:28.41 & 0.9748 & 4 \\
14& 14:21:18.842 & 53:00:32.05 &  0.7395 & 4 \\
15& 14:21:21.818 & 53:01:55.12 & 0.7321 & 4 \\
\hline
\end{tabular}
\end{center}
\label{tab:sample1}
\end{table}

\begin{table}[t!]
  \caption{Running number, coordinates (J2000), redshifts, and spectral flags for the objects observed 
    with GEMINI/GMOS in the D3-6 field of view.}
\begin{center}
\begin{tabular}{rrrrr}
\hline
\hline
\#  & RA           &  DEC       & redshift & flag \\
  \hline         
1 & 14:16:36.004 & 53:04:31.73 & 0.7163 & 3 \\
2 & 14:16:37.152 & 53:04:32.32 & 0.5268 & 3 \\
3 & 14:16:38.064 & 53:03:44.66 & 0.7441 & 2 \\
4 & 14:16:39.084 & 53:05:44.23 & 0.6083 & 4 \\
5 & 14:16:40.173 & 53:05:01.37 & 0. & 4 \\
6 & 14:16:41.339 & 53:04:54.53 & 0.6474 & 3 \\
7 & 14:16:46.077 & 53:01:47.67 & 0.5758 & 4 \\
8 & 14:16:46.824 & 53:04:38.55 & 0.6090 & 4 \\
9 & 14:16:47.833 & 53:03:04.95 & 0.6046 & 4 \\
10& 14:16:49.100 & 53:05:55.98 & 0.6067 & 3 \\
11& 14:16:50.122 & 53:05:10.06 & 0.6077 & 3 \\
12& 14:16:51.068 & 53:02:17.15 & 0.6007 & 2 \\
13& 14:16:53.373 & 53:05:51.77 & 0. & 4 \\
14& 14:16:55.542 & 53:05:16.50 & 0.6064 & 4 \\
15& 14:16:56.983 & 53:02:43.27 & 0.9006 & 3 \\
16& 14:16:57.645 & 53:04:29.12 & 0.6074 & 2 \\
17& 14:16:58.438 & 53:03:47.13 & 0.5888 & 2 \\
18& 14:16:59.666 & 53:02:26.51 & 0.6064 & 4 \\
19& 14:17:00.703 & 53:05:15.49 & 0.9670 & 4 \\
20& 14:17:01.736 & 53:03:19.29 & 0.6441 & 3 \\
21& 14:17:03.493 & 53:00:50.72 & 0.5255 & 2 \\
22& 14:17:04.775 & 53:03:15.64 & 0.7748 & 4 \\
23& 14:17:05.915 & 53:05:58.64 & 0.5255 & 3 \\
24& 14:17:06.757 & 53:05:55.87 & 0.5270 & 4 \\
25& 14:17:08.225 & 53:00:47.94 & 0.6478 & 4 \\
26& 14:17:09.578 & 53:03:34.14 & 0. & 4 \\
\hline
\end{tabular}
\end{center}
\label{tab:sample2}
\end{table}

\subsection{Publicly available spectroscopy}

One of the two studied candidate clusters (D3-43) was also covered by the KECK/DEIMOS
DEEP2 spectroscopic survey (Newman et al. 2013). 
This provided 336 redshifts in a 5~arcmin radius
circle around the cluster centre, in addition to the GEMINI/GMOS
redshifts.  Ten of these GEMINI/GMOS redshifts were also measured
by the KECK/DEIMOS DEEP2 survey, and this allowed us to assess our own
redshift measurements, as shown in Fig.~\ref{fig:zz}. We do not detect any
noticeable difference. Even the two galaxies with the most different
redshift values in the two surveys do not exhibit a redshift
difference larger than 0.015. This typically corresponds to the
3$\sigma$ uncertainty according to the smoothing we applied to
GEMINI/GMOS spectra before redshift measurements (dashed lines in
Fig.~\ref{fig:zz}).
If we consider only redshifts lower than 0.8, the typical uncertainty
between the two redshift measurements is 0.0005.

The other candidate cluster (D3-6) has only three additional objects
with known spectroscopic redshifts available through the NED and
Simbad databases (Howell et al. 2005, Hsieh et al. 2005, Walker et
al. 2011). None are in common with our GEMINI/GMOS catalogue.

\subsection{The main structures along the two lines of sight}

\begin{figure}[!ht]
  \begin{center}
    \includegraphics[angle=270,width=3.5in]{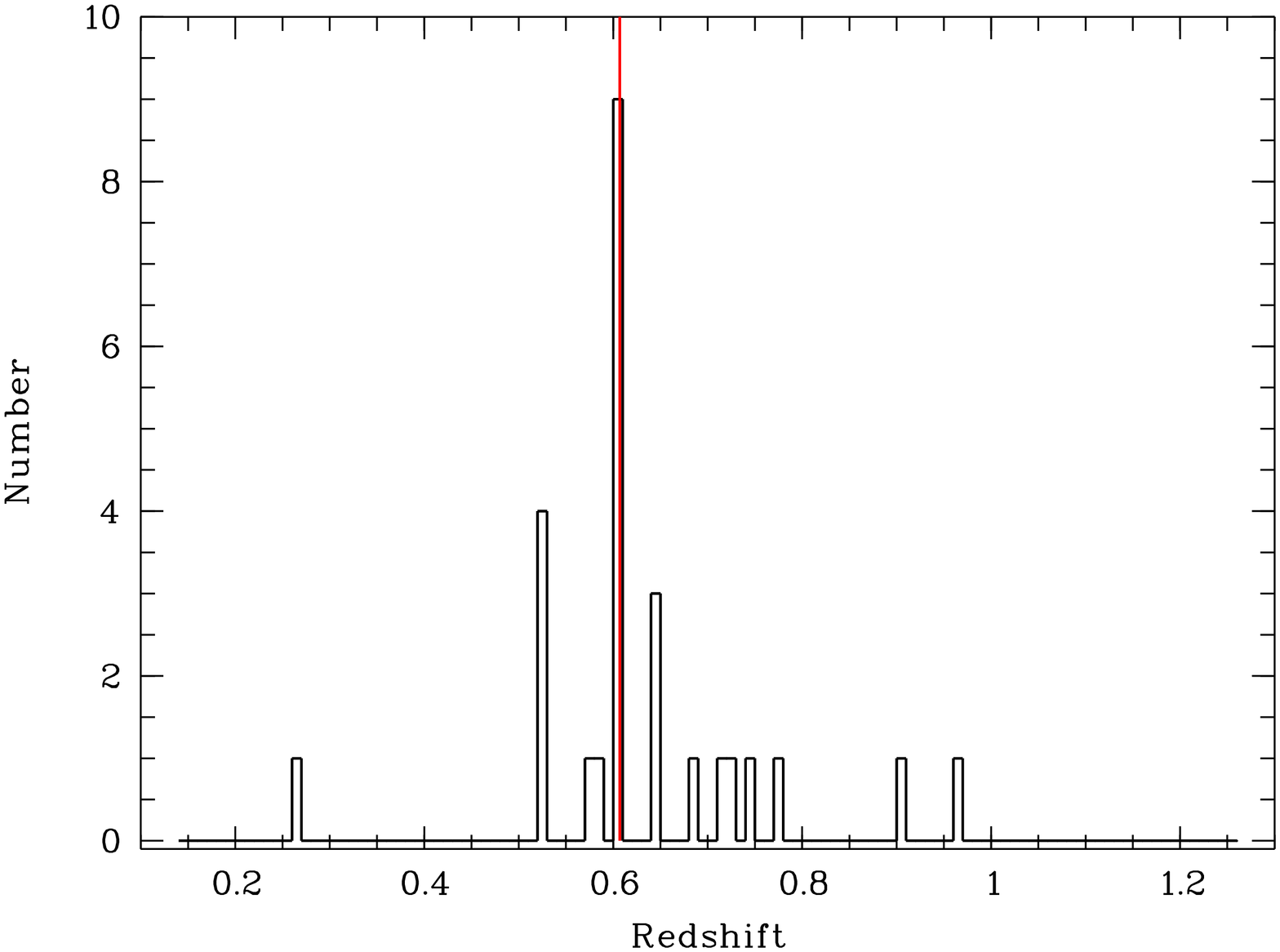}
    \includegraphics[angle=270,width=3.5in]{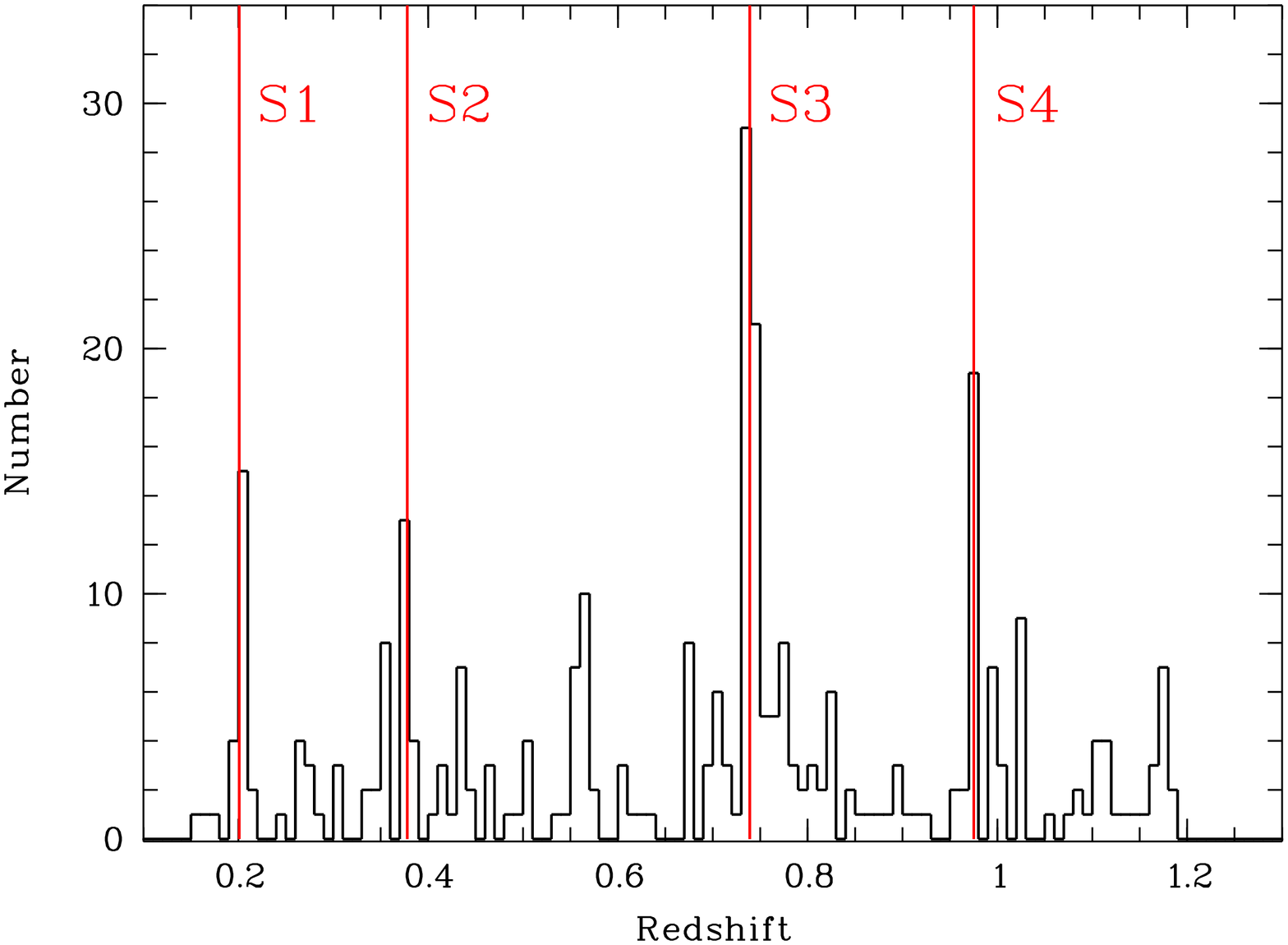}
    \caption{Distribution of the Gemini/GMOS and DEEP2 galaxies with known spectroscopic
      redshifts along the D3-6 (top), and D3-43 (bottom) lines of
      sight. Detected structures from Table~3 are shown as red vertical
      lines.}
  \label{fig:bighisto}
  \end{center}
\end{figure}

We first merged our own redshift catalogue with those of the literature
and eliminated the galaxies observed twice, as well as the stars. We
then computed redshift histograms along the two lines of sight, and
searched for empty gaps wider than 0.001 in the redshift
distribution. This is a classical method to detect galaxy structure
candidates (e.g. Katgert et al. 1996). We only retained galaxy
concentrations between two such gaps of more than five objects.  This
provided four potential structures along the D3-43 line of sight and
only one along the D3-6 line of sight (see Table~3 and
Fig.~\ref{fig:bighisto}).

\begin{figure}[!ht]
  \begin{center}
    \includegraphics[angle=270,width=3.5in]{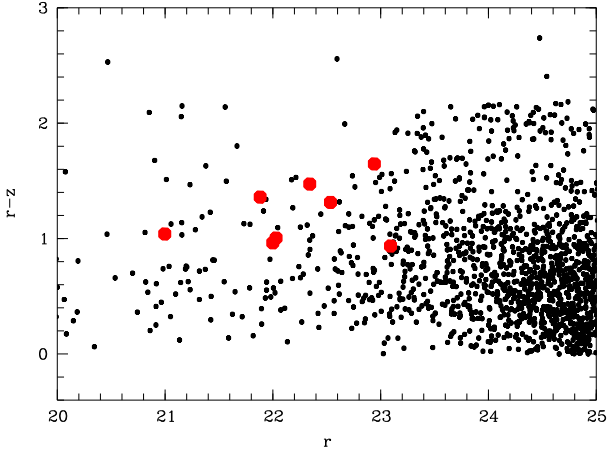}
    \caption{Colour-magnitude relation along the D3-6 cluster line of
      sight. Small black dots are the CFHTLS galaxies, red disks are
      the spectroscopically confirmed cluster members.}
  \label{fig:cmr6}
  \end{center}
\end{figure}

The D3-6 line of sight only shows one structure at z=0.607 (see
Fig.~\ref{fig:cmr6}) sampled with height galaxies with spectroscopic
redshifts. The $(r^\prime - z^\prime)$ values of these galaxies are
relatively similar and of the order of 1.4 and 1, also in good
agreement with the colours of z=0.5 early-type and spiral galaxies
respectively (e.g. Fukugita et al. 1995).  The mean redshift of the
only detected structure along the D3-6 line of sight (z=0.607) also
being close to the candidate cluster photometric redshift value
(z=0.6), we decided to associate it with the D3-6 candidate cluster of
Adami et al. (2010).

\begin{table}
  \caption{Structures detected along the line of sight to the two candidate clusters.
    The columns are: (1)~cluster name, 
    (2)~number of the structure,
    (3)~number of galaxies in the structure, (4)~mean redshift of the
structure, (5)~Rostat velocity dispersion,
    (6)~number of substructures when analysed by the Serna \& Gerbal (1996) method. }
\begin{tabular}{lrrrrr}
\hline
\hline
Name & \#  & N$_{gal}$ & $z$~~ & vel. disp. (km/s ) & SG substructures \\
\hline
D3-6              & 1 & 8  & 0.607 & 423  & 1 \\ 
\hline
D3-43             & S1 & 17 & 0.201 & 489  & / \\        
                  & S2 & 11 & 0.378 & 250  & / \\        
                  & S3 & 46 & 0.739 & 1152 & 2 \\        
                  & S4 & 18 & 0.975 & 575  & / \\        
\hline
\end{tabular}
\label{tab:los}
\end{table}

\begin{figure}[!ht]
  \begin{center}
    \includegraphics[angle=270,width=3.5in]{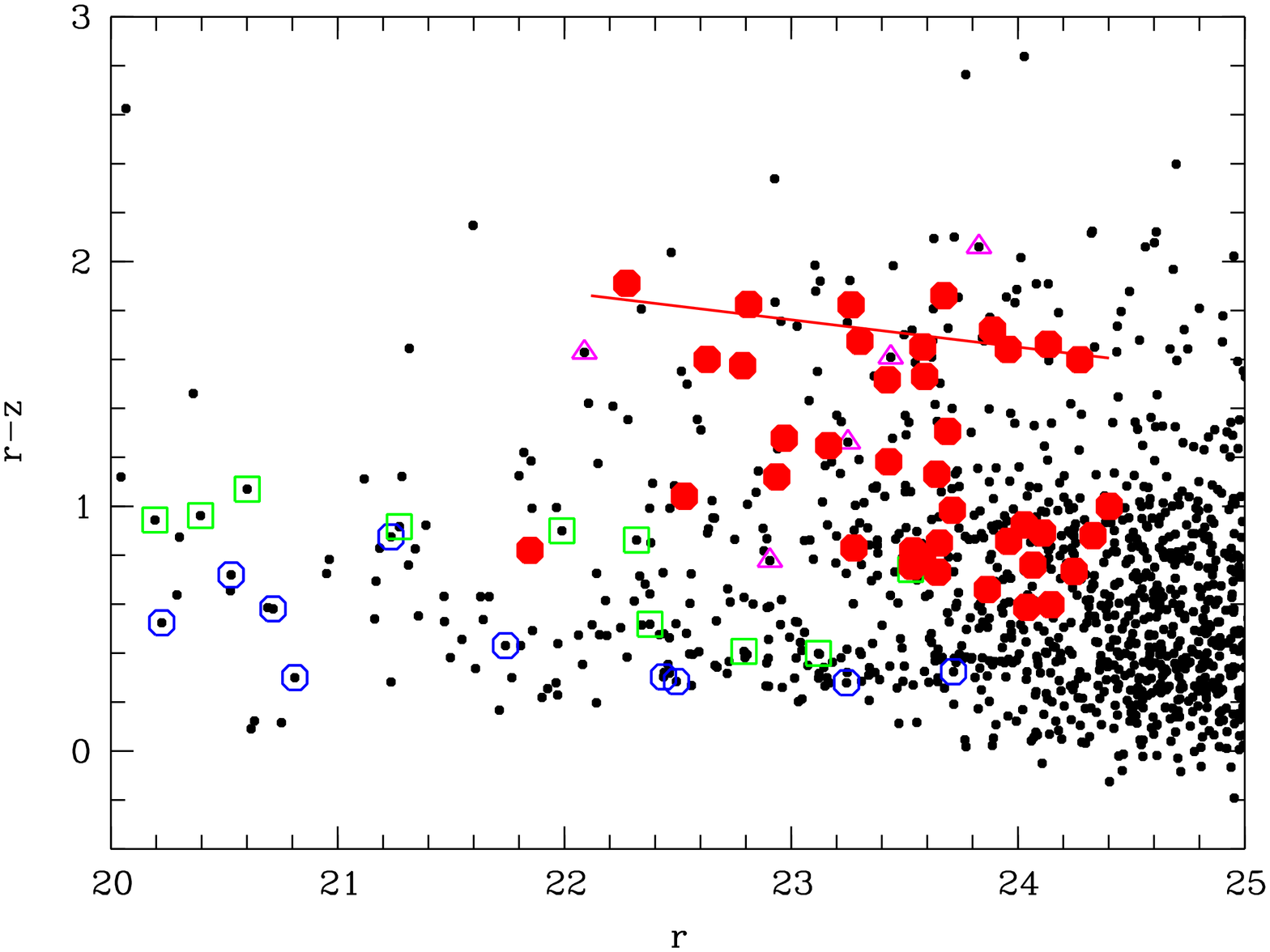}
    \includegraphics[angle=270,width=3.5in]{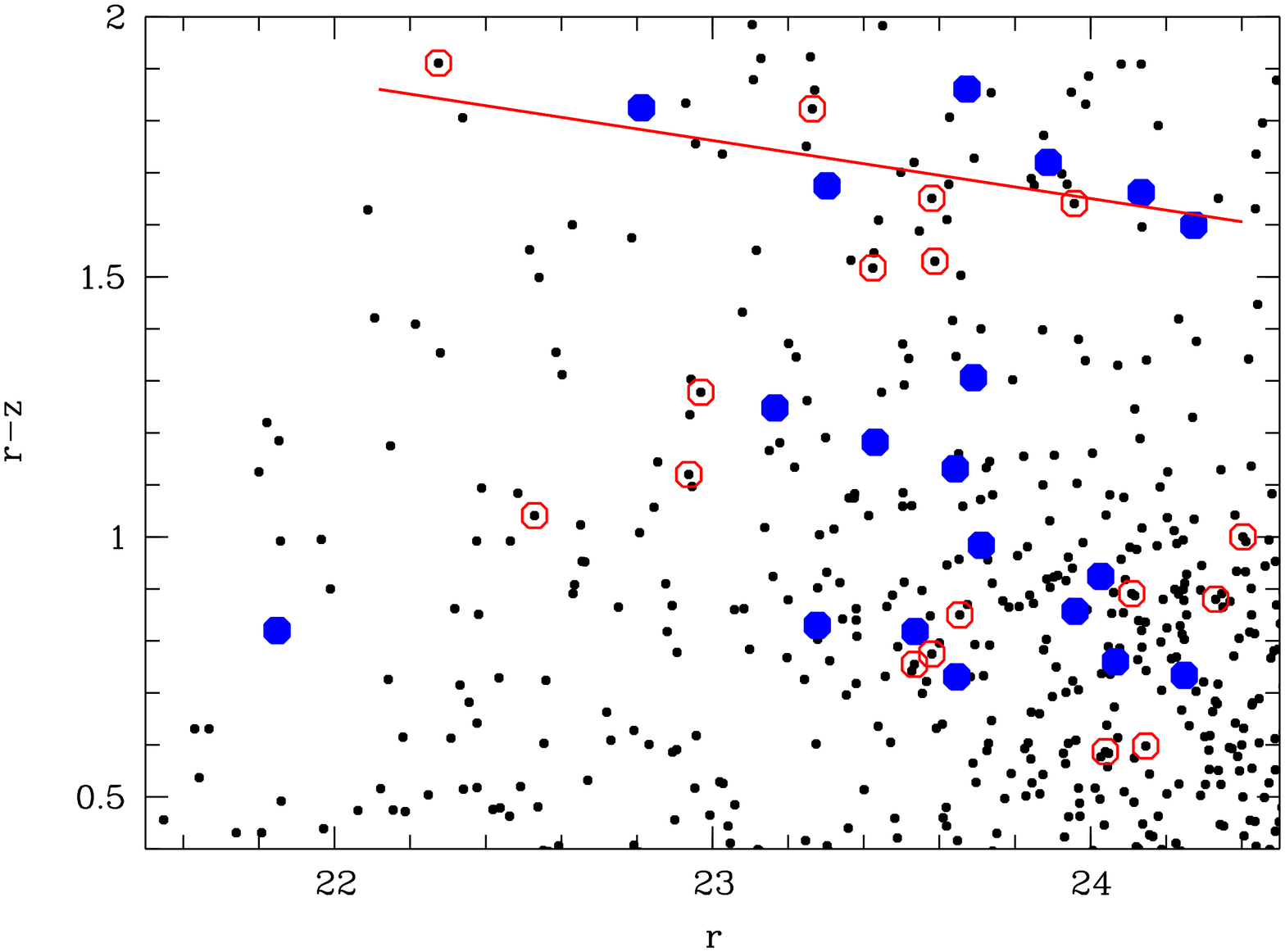}
    \caption{Colour-magnitude relations along the D3-43 candidate cluster line of
      sight. Small black dots are the CFHTLS galaxies. Upper figure:
      red disks are the spectroscopically confirmed members of
      S3, blue circles are the S1 members, green squares are the S2
      members, pink triangles are the S4 members. The inclined red line 
      is the tentative red sequence for structure S3 early-type galaxies.
      Lower figure: red circles are the S3 substructure \#1 members, blue disks are the S3
      substructure \#2 members (see text for details). The inclined red line 
      is the tentative red sequence for S3 early-type galaxies.}
  \label{fig:cmr43}
  \end{center}
\end{figure}

Fig.~\ref{fig:cmr43} displays the colour-magnitude relations of the
potential structures along the D3-43 line of sight in the $(r^\prime -  z^\prime)$
versus $r^\prime$ space.  We clearly see a red sequence (showing an old galaxy
population) around $(r^\prime -  z^\prime)$$\sim$1.8 and a bluer cloud around
$(r^\prime - z^\prime)$$\sim$1 for structure S3. These colors are consistent
with z=0.8 early and late-type galaxies (e.g. Fukugita et
al. 1995).  Other structures of the D3-43 line of sight are less
prominent and do not show the same dichotomy between an early-type and
a late-type galaxy population. The structure S3 is also by far the most populated:
it has more confirmed galaxies than the three others taken
together. The structure S3 is clearly the closest to the 3$\sigma$ peak of the weak
lensing detection (see below and Fig.~\ref{fig:D3-43WL}). The
brightest galaxy of S3 is at $\sim$600 kpc of the peak, while the
brightest galaxies of S1, S2, and S4 are located at $\sim$850 kpc,
$\sim$2300 kpc, and $\sim$1300 kpc (at the structure redshift). The structure S3 is
the only structure for which the galaxy dispersion on the sky is
larger than the distance between the structure centre and the
3$\sigma$ peak of the weak lensing detection (1.2 $\times$
larger). The ratio is 0.4 for S1, 0.3 for S2, and 0.9 for S4.
Finally, the spectroscopic redshift of S3 is remarkably close to the
photometric redshift of D3-43 in Adami et al. 2010: 0.75.  We
therefore decided to choose S3 as the main structure along the D3-43
line of sight, to associate it with the weak lensing detection (see
below), and with the D3-43 candidate cluster of Adami et al. (2010).

\begin{figure}[!ht]
  \begin{center}
    \includegraphics[angle=270,width=3.5in]{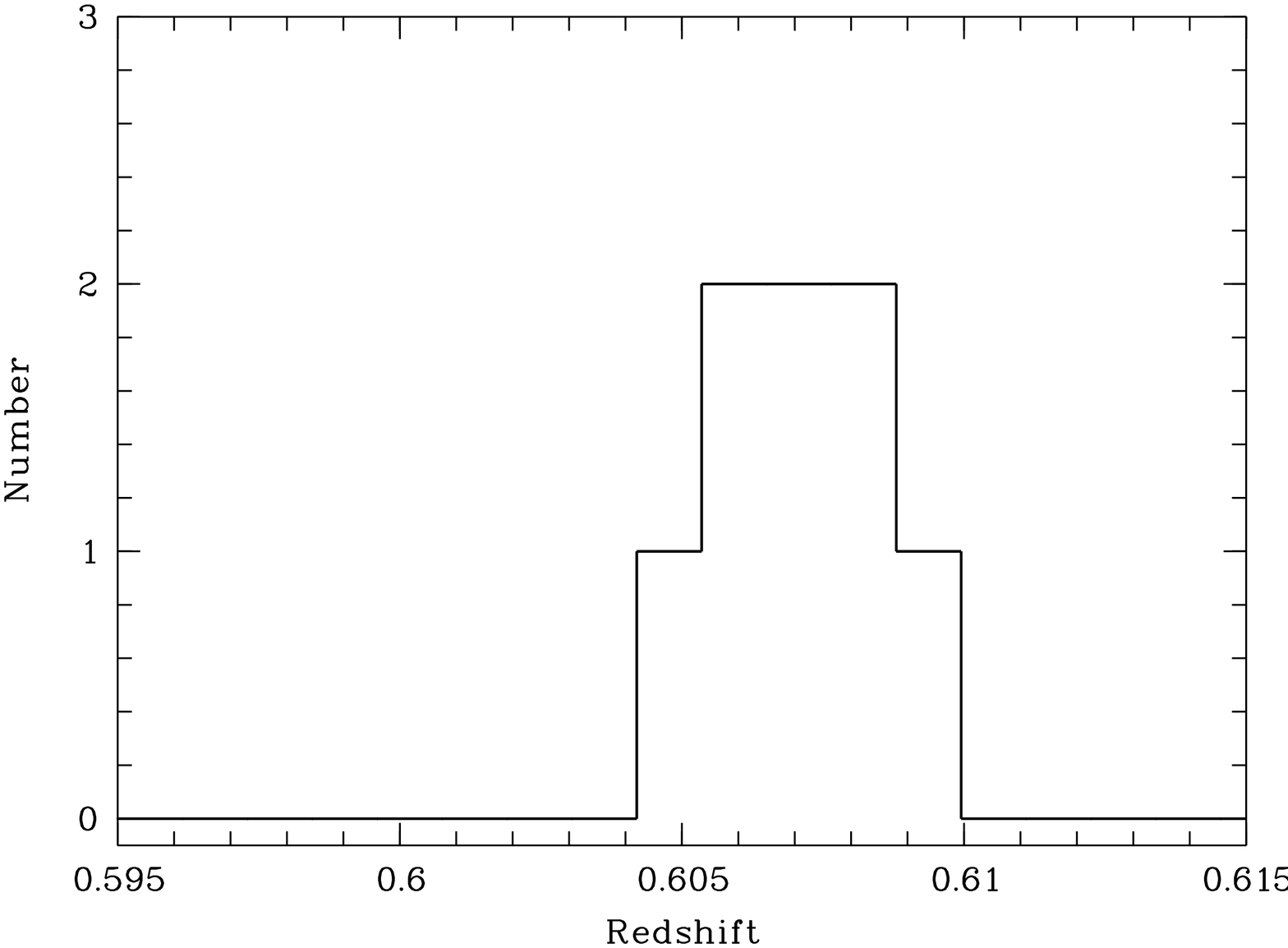}
    \includegraphics[angle=270,width=3.5in]{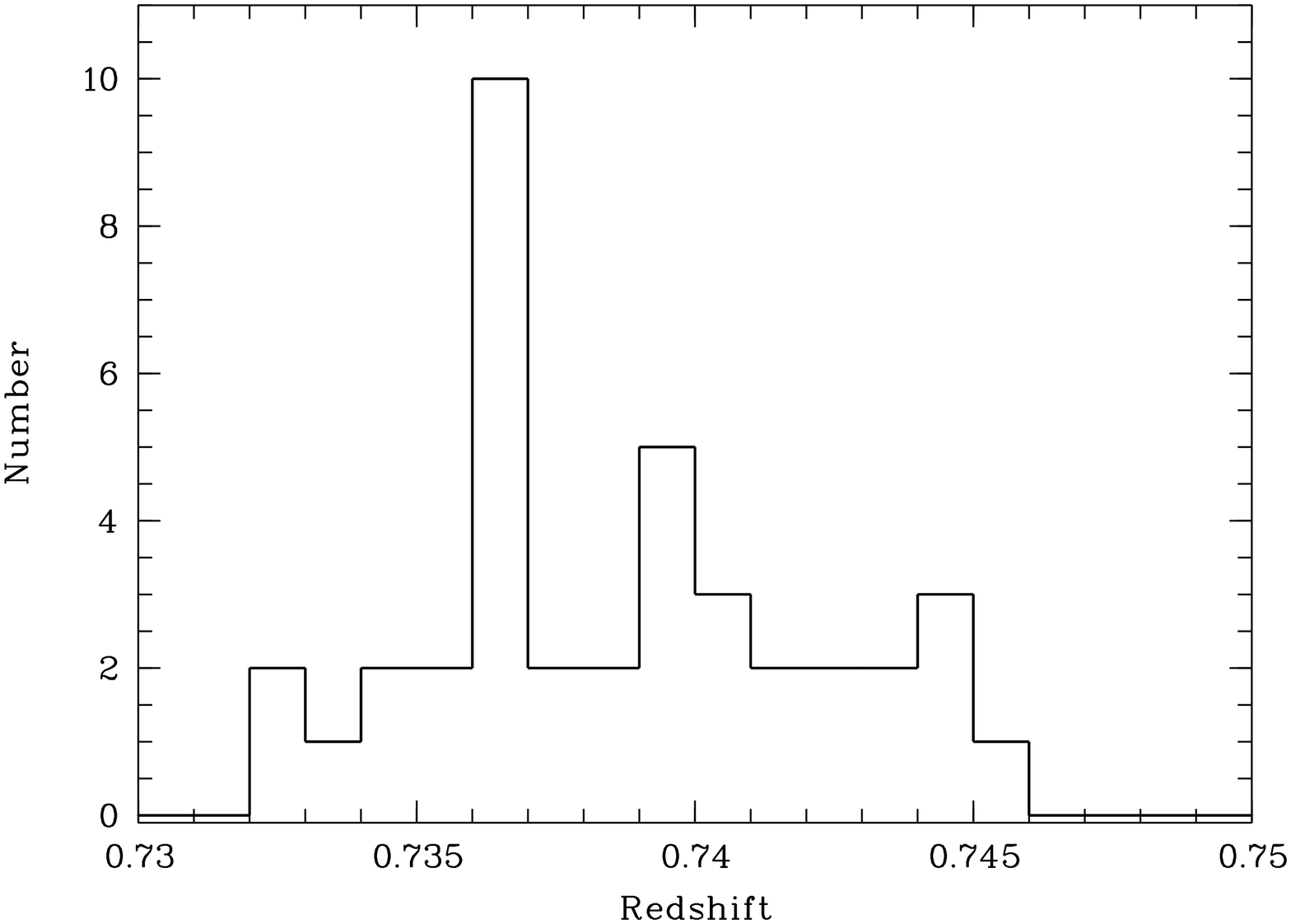}
    \caption{Redshift distribution of the Gemini/GMOS galaxies with
      known spectroscopic redshifts in the D3-6 (top), and D3-43-S3
      (bottom) structures.}
  \label{fig:histoz}
  \end{center}
\end{figure}

This leads us to conclude that we probably have detected at least two
real structures along the D3-6 and D3-43 lines of sight 
at z=0.607 (D3-6 in the following) and z=0.739 (D3-43-S3 in the
following) respectively. The redshift distributions of the galaxies with known
spectroscopic redshifts inside these structures are displayed in
Fig.~\ref{fig:histoz}.

\subsection{Internal structure of D3-6 and D3-43-S3}

We investigated the internal structures of D3-6 and D3-43-S3 by applying the Serna-Gerbal
technique (hereafter SG, 1996 release, Serna \& Gerbal 1996). This hierarchical code
based on spectroscopic redshifts and optical magnitudes, is designed
to detect substructures in the optical.  A number of other methods are
available to search for substructures at optical wavelengths, such as the
$\Delta -$test (Dressler \& Schechtman 1988). However, the SG method
has proven to be quite powerful detecting evidence for substructuring in
nearby (see Abell~496: Durret et al. 2000; Coma: Adami et al. 2005,
2009; Abell~780: Durret et al. 2009; Abell~85: Bou\'e et al. 2008),
moderate redshift (Abell~222/223: Durret et al. 2010), and high
redshift clusters (RX~J1257.2+4738: Ulmer et al. 2009). Recently, it
was also successfully applied and compared with X-ray detections on a
larger sample (the DAFT/FADAS sample: Guennou et al. 2014).  Assuming
a value of the mass to luminosity ratio (here taken to be 100 in the 
$r^\prime$ band, to be homogeneous with the Guennou et al. 2014 simulations), the SG
method allows us to estimate the masses of the substructures that it
detects.  Guennou et al. (2014) have shown that although the absolute
masses are not accurate (typical uncertainties are clearly larger than $10^{14}$ 
M$_{\odot}$), the mass ratios of the various substructures
were well determined.  The SG method has also been extensively tested
on simulations by Guennou (2012), in particular concerning the effect
of undersampling on mass determinations. We stress that this method
requires a very good precision in the galaxy distance determinations,
and therefore spectroscopic redshifts are essential, while photometric
redshifts are inappropriate.

The level of refinement in the substructure detection obviously
depends on the spectroscopic sampling, as already shown by the
previously quoted articles. We first processed the D3-6
structure. Since it is only sampled by eight galaxies with known spectroscopic redshifts, the
SG code would only be able to detect large substructures that would be
present in cluster-cluster merging, and this is not the case, showing
that D3-6 is probably not undergoing a major merger. 

The D3-43-S3 structure is much better sampled spectroscopically. The SG method 
detects two relatively massive substructures (see Table 4) and several other more 
dynamically isolated galaxies.  Substructure
\#1 includes the dominant galaxy of D3-43-S3 and has an
estimated dynamical mass of $3.6 \times 10^{14}$~M$_{\odot}$. The estimated
mass of substructure \#2 is lower, of the order of $8.8 \times
10^{13}$~M$_{\odot}$. D3-43-S3 is therefore probably about to undergo an
important merger between at least two comparable galaxy structures (see also 
section 5). To confirm this statement, we also show in Fig.~\ref{fig:cmr43}
(bottom part) the colour-magnitude relation of the two substructures
in the D3-43-S3 region. This
shows that the two substructures have both red and blue galaxies. We
are therefore not dealing with a structure populated by young galaxies
merging with an older galaxy structure. The apparently high velocity dispersion of
D3-43-S3 (when not considering the detected substructures) is also consistent with this scenario.

\begin{table*}[t!]
  \caption{Main characteristics of the two confirmed structures. The columns are: 
    (1) name, 
    (2) + (3) coordinates in the D3 CFHTLS cluster catalogue, 
    (4) photometric redshift in the D3  CFHTLS cluster catalogue, 
    (5) and (6) central coordinates of the spectroscopic catalogue defined 
    as the coordinates of the brightest galaxy member, 
    (7) mean spectroscopic redshift, 
    (8) velocity dispersion from Rostat estimates (Beers et al. 1991) (values are also given for the two substructures
of D3-43-S3),
    (9) mass estimate from the Serna-Gerbal analysis (values are also given for the two substructures
of D3-43-S3),
    (10) mass estimate from the X-ray analysis (within r$_{500}$). }
\begin{tabular}{lrrrrrrccc}
\hline
\hline
Name         &  RA(D3)    & DEC(D3)  & z(D3) & RA(spec) & DEC(spec) & z(spec) & vel. disp. & SG Mass   & X-ray Mass \\
             &  deg       & deg      &       & deg      & deg       &         & km/s       & $M_{\odot}$ & $M_{\odot}$ \\
\hline
D3-6         &  214.2207 & 53.0568   & 0.60 & 214.2314  & 53.0879   &  0.607    &  423  & $2.2 \times 10^{14}$ &  $2.5 \times 10^{14}$\\
D3-43-S3     &  215.271 & 53.0353   & 0.75 & 215.2710  & 53.0353   &  0.739    &  1152 (611 + 357) & $3.6 \times 10^{14} + 8.8 \times 10^{13}$ & $\leq 8 \times 10^{13}$ \\ 
\hline
\end{tabular}
\label{tab:mainchar}
\end{table*}

\section{Mass characterisation of the D3-6 and D3-43-S3 structures}

At this stage, we have secured the detection of the D3-6 and D3-43-S3
structures, confirmed that they are populated by both red and blue
galaxies, shown that D3-6 is not undergoing a major merger, and 
that D3-43-S3 is probably about to undergo such a merger. However, we do not yet have
robust estimates of the masses of these two structures (the SG test does
not give reliable absolute mass values, as already stated).

\subsection{Publicly available X-ray data}

Both fields D3-6 and D3-43-S3 were observed by the \textit{Chandra}
X-ray telescope. Even if the collecting area of \textit{Chandra} is
not favourable to the characterisation of such distant structures, the
exposure times were long enough to at least give an
estimate of the X-ray luminosities of these structures.

There are four pointings at the D3-43 region with the ACIS-I detector
(obs\_id 5845, 5846, 56214, 6215, PI K. Nandra, taken in
2005). Following the ``Science Threads'' from the Chandra X-ray Center
(CXC), using \textsc{ciao}
4.6\footnote{\texttt{asc.harvard.edu/ciao/}}, we have reprocessed
these observations and merged them together producing a single
broad-band (0.5--7.0 keV) exposure-map corrected surface brightness
image with a pixel scale of $1.968$~arcsec (14/$h_{70}$ kpc assuming
$z=0.739$). The total effective exposure time for this image is
193.6~ks.  There is apparently no visible large scale extended X-ray
emission coinciding with the position of D3-43-S3. We only see a
collection of three compact sources. In a circular region of radius $R
= 2$ arcmin, we estimate that an extended source with count rate above
$4.2 \times 10^{-4}\,$cnt/s (about 85 counts) would be detected. Thus,
assuming a plasma thermal emission, we can put an upper limit of $f_X
< 4.3 \times 10^{-15}\,$erg~s$^{-1}$~cm$^{-2}$. This corresponds to an
upper limit for the X-ray luminosity $L_X < 1.5 \times
10^{43}\,$erg~s$^{-1}$ if the source is at $z = 0.739$. This could be
typical of a massive structure of galaxies, but at this stage this is not 
enough to reach a conclusion about the massive nature of D3-43-S3.

\begin{figure}[!ht]
  \begin{center}
    \includegraphics[angle=0,width=3.5in]{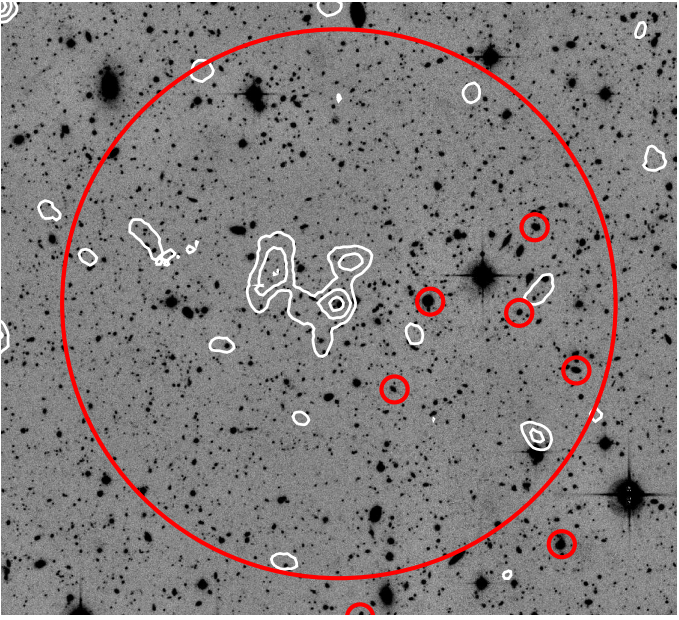}
    \caption{CFHTLS $i^\prime$ band image of D3-6 overlaid with
      \textit{Chandra} contours in white (see text). Small red circles
      show the galaxies with known spectroscopic redshifts inside the
      z=0.607 D3-6 structure. The large red circle represents a 1~Mpc
      radius area.}
  \label{fig:D3-6X}
  \end{center}
\end{figure}

The D3-6 region was observed in April 2002 with ACIS-I (obs\_id 3239,
P.I. E.~Ellingson) with an exposure time of 62.82~ks. We followed the
same reduction procedure, producing a flat (exposure-corrected) image
in the 0.5--7.0 keV band with a pixel scale of $1.968$~arcsec
(13/$h_{70}$ kpc assuming $z=0.607$). A faint, extended source is
visible at 14:17:01.7, +53:05:13 (J2000).  To produce an
image showing only the diffuse component, we have detected all point
sources with the task \textsc{wavdetect}, following \textsc{ciao}
Science Threads\footnote{http://cxc.harvard.edu/ciao/threads/}, which
is based on the wavelet image decomposition technique. Then, the
regions containing the detected point sources were replaced with a
Poissonian noise, using the \textsc{dmfilth} task, with the same mean
value sampled from an elliptical annulus around the source. Finally,
for display purposes, we have smoothed the diffuse emission image
with a Gaussian kernel of 12 pixels (23.6$^{\prime\prime}$). Within
2~arcmin, we estimate a net count rate (background subtracted) of
$(5.56 \pm 0.97) \times 10^{-3}$~cnt/s. This corresponds to a flux
$f_X = (5.44\pm 0.92) \times 10^{-14}$~erg~s$^{-1}$~cm$^{-2}$ and,
assuming a redshift $z=0.607$, a bolometric luminosity $L_X = (1.12
\pm 0.19) \times 10^{44}$~erg~s$^{-1}$, typical of a low-mass cluster
of galaxies.  The derived X-ray luminosity corresponds to a velocity
dispersion $\sigma _v \approx 500$~km~s$^{-1}$, using the scaling
relation by Lopes et al. (2009). This estimate is very close to our
Serna-Gerbal value (see Table 4).  We show in Fig.~\ref{fig:D3-6X} the
\textit{Chandra} contours and the galaxies with known spectroscopic redshifts in the
D3-6 cluster overlaid onto the CFHTLS $i^\prime$ band image. Five of the
galaxies with known spectroscopic redshifts are at less than 1 Mpc from the cluster X-ray
centre, and this is probably enough to secure the association between
the X-ray emission and the galaxy redshift concentration (see
e.g. Adami et al. 2011 for the XMM-LSS survey). This association would
need to be definitively confirmed by measuring the redshift of
the X-ray central galaxy, however.

If we convert the X-ray luminosities to masses by applying the scaling
relation of Lopes et al. (2009), we find a mass of $2.5\times 10^{14}$~M$_\odot$ for
D3-6, and an upper limit of $8\times 10^{13}$~M$_\odot$ for D3-43
(calculated within $r_{500}$).

\subsection{Weak lensing characterisation of D3-43-S3}

To check if we can investigate the mass distribution around
D3-43-S3 in an independent way, we took advantage of the CFHTLS D3
images in the $i^\prime$ band for which a special release is available with a
seeing of 0.64~arcsec. Even with such a good ground-based seeing, it
is still a difficult task to try to detect modest mass concentrations
at z$\sim$0.74 , but we have no other possibility with the data available
to tell if D3-43-S3 has a significant mass or not.

\begin{figure}[!ht]
  \begin{center}
    \includegraphics[angle=0,width=3.5in]{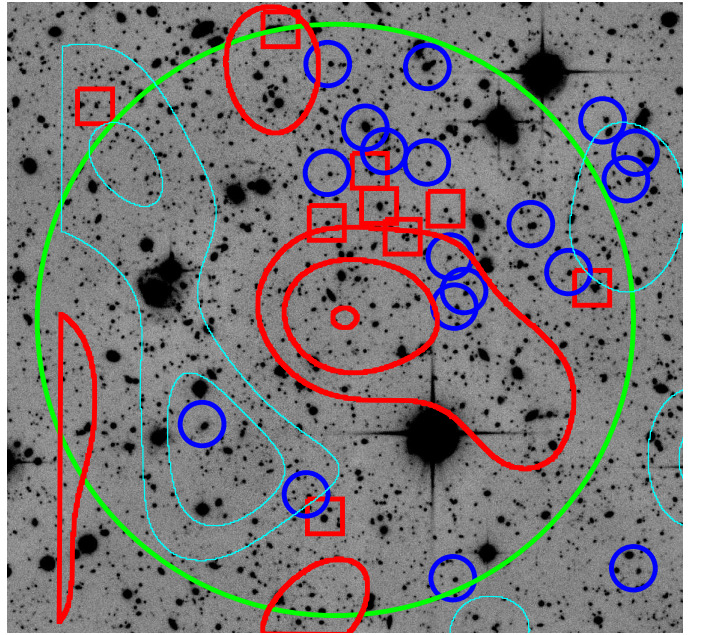}
    \caption{CFHTLS $i^\prime$ band image of D3-43 overlaid with weak lensing
      contours (see text). Red squares and  blue circles
      are the galaxies with known spectroscopic redshifts inside
      the z=0.739 D3-43-S3 substructures \#1 and \#2, respectively. The
      green circle has a radius of 1~Mpc. We show the positive
        sources as thick red contours and the negative sources as thin
        cyan contours. The contours are given in numbers of sigma
        corresponding to the noise in the convergence map
        reconstruction, starting with $1\sigma$.}
  \label{fig:D3-43WL}
  \end{center}
\end{figure}

We cut a subimage centred on our structure with a field of view of
6.3'x6.3' (corresponding to $2.74 \times 2.74$~Mpc$^2$). This is
roughly the size of VLT/FORS2 images, which are well suited for cluster weak
lensing studies at this redshift (e.g. Clowe et al. 2006). We measure
object positions using SExtractor (Bertin \& Arnouts 1996) and shapes
using the latest {\it imcat} software tools (Kaiser 2011). We then
apply the standard KSB+ methodology for PSF correction (Kaiser et
al. 1995, Luppino \& Kaiser 1997). The full detail of the method
applied will be found in Martinet et al. (2015, in preparation). The
general idea is to measure the PSF distortion on the stars in the
image to subtract it from the galaxy shape measurements. In
practice, this information is retained in the fourth moment of object
surface brightness distributions. Stars are discriminated from
galaxies in a half-light radius versus magnitude plot. A visual
inspection of both stars and galaxies in the field is mandatory to
eliminate false detections, objects that are blended or near
saturated stars and artefacts. We also apply a correction factor
that represents the bias of our method and was calibrated on STEP2
simulations (Massey et al. 2007). We eliminate cluster and foreground
galaxies to avoid diluting the signal, by considering photometric
redshifts computed on the basis of $u^*$, $g^\prime$, $r^\prime$, $i^\prime$, $z^\prime$, J, H, Ks
(CFHTLS and WIRDS release, see Bielby et al. 2012) magnitudes. Given
the precision of these photometric redshifts, we remove all objects at
redshifts lower than the structure redshift plus 0.1, keeping only
distant galaxies, the only ones potentially sensitive to the D3-43-S3
mass concentration. Finally, the shear catalogue is converted into a binned 
shear map, which is then inverted into a convergence map following Kaiser $\&$ Squires (1993). 
We then apply a Gaussian smoothing that allows us to estimate the noise level in the map (Van 
Waerbeke 2000). The density of background galaxies is $\sim18$ galaxies per square arcmin. 
The contours are given in numbers of sigma corresponding to the noise in the convergence map 
reconstruction. The result is that D3-43-S3 is barely detected at a 3$\sigma$ level (see
Fig.~\ref{fig:D3-43WL}).  We note that given the quite low detection
level, we did not try to calibrate the weak lensing countours in terms
of mass.
 
We show in Fig.~\ref{fig:D3-43WL} the weak lensing contours and the
galaxies with known spectroscopic redshifts in D3-43-S3 overlaid on the CFHTLS $i^\prime$
band image. Twentytwo of the galaxies with known spectroscopic redshifts are at less than
1 Mpc from the cluster weak lensing centre.
We will not discuss the weak lensing analysis of D3-6 since its detection level
is smaller than 3$\sigma$.

\section{Properties of the galaxy populations in the two structures}

Publicly available photometric redshifts in the CFHTLS D3 field are
among the best available from ground-based data (computed with
LePhare, e.g. Coupon et al. 2009). Following Coupon et al. (2009), we chose to 
limit our photometric redshift samples to $i^\prime$$\leq$24 ($g^\prime$$\leq$25 with
Fukugita et al. 1995).

We also verified that for the spectroscopic member galaxies of D3-43-S3,
the statistical uncertainty of the photometric redshifts was of the
order of 0.037. To keep as many structure galaxies 
as possible in our samples, we then chose to select a slice
of $\pm$3$\times$0.037 around the two clusters. The price to pay is
the likely inclusion of field galaxies satisfying this criterium. This
contribution is quite easy to remove statistically however, using for
example a comparison field (CF hereafter) taken in the CFHTLS D3 field
(with the same angular size as the considered cluster), empty
from any known candidate cluster and with galaxies selected in the same
photometric redshift range.

The remaining uncertainty in the cluster galaxy counts comes from the
catastrophic errors present in the photometric redshifts. To take this
uncertainty into account, we considered the percentage of such
catastrophic errors in the CFHTLS deep fields computed in Coupon et
al. (2009): 4$\%$. This was added to the Poissonian error bars
estimated for each magnitude bin in what follows.

We then re-ran the LePhare code with both the Cosmos-survey SEDs
(e.g. Ilbert et al. 2010) and the Bruzual $\&$ Charlot (2003) SEDs,
fixing the redshifts of the selected samples to the spectroscopic
redshifts of the two structures. We only considered galaxies for which
all the bands were available.  This gave us rest-frame absolute
magnitudes, stellar masses, and star formation rates, hereafter SFR.

\subsection{Galaxy luminosity functions}

Galaxy luminosity functions were computed by counting the galaxies along
the D3-6 and D3-43-S3 lines of sight and statistically subtracting the empty
field counts.

\begin{figure}[!ht]
  \begin{center}
    \includegraphics[angle=270,width=3.5in]{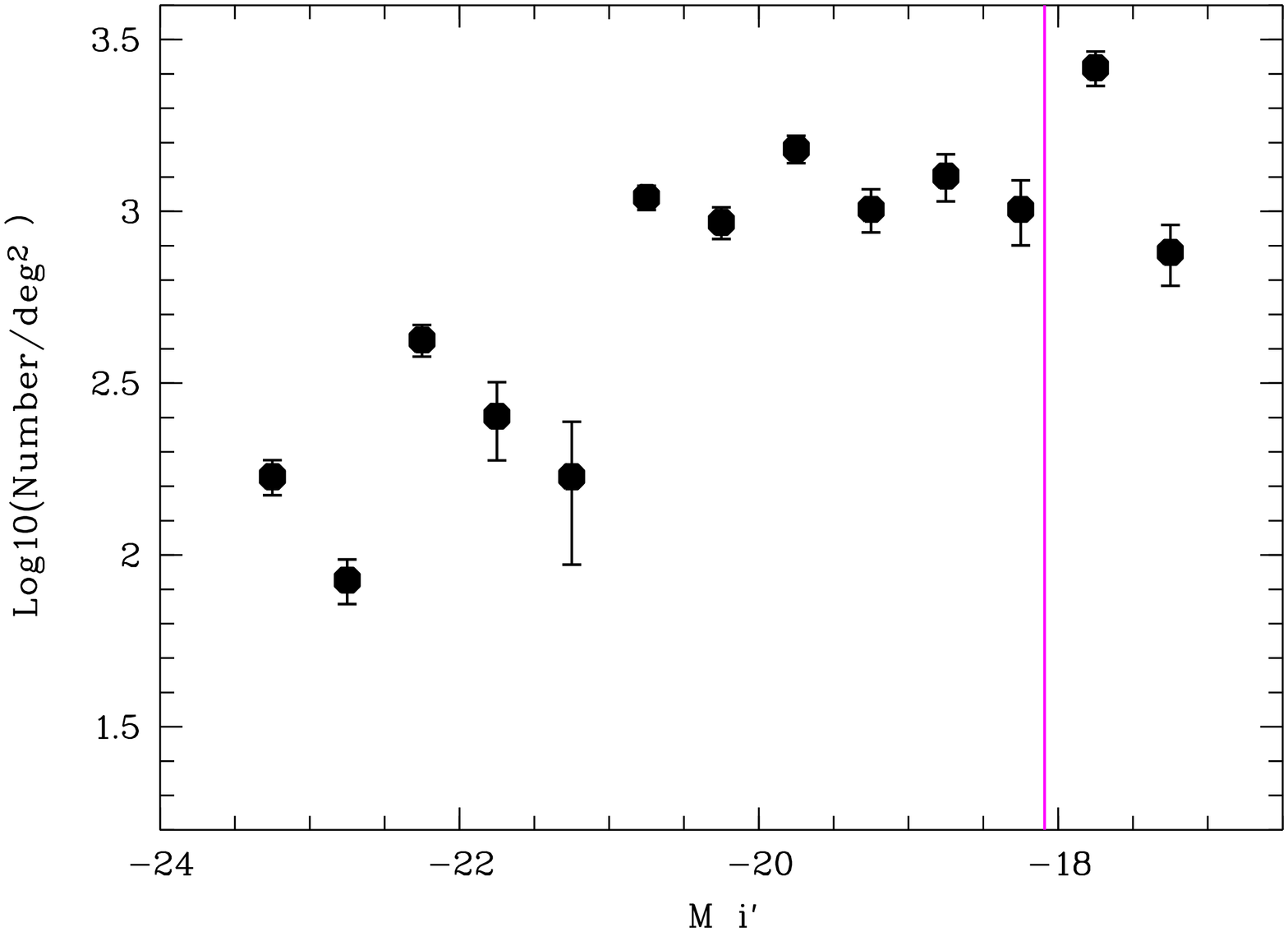}
    \includegraphics[angle=270,width=3.5in]{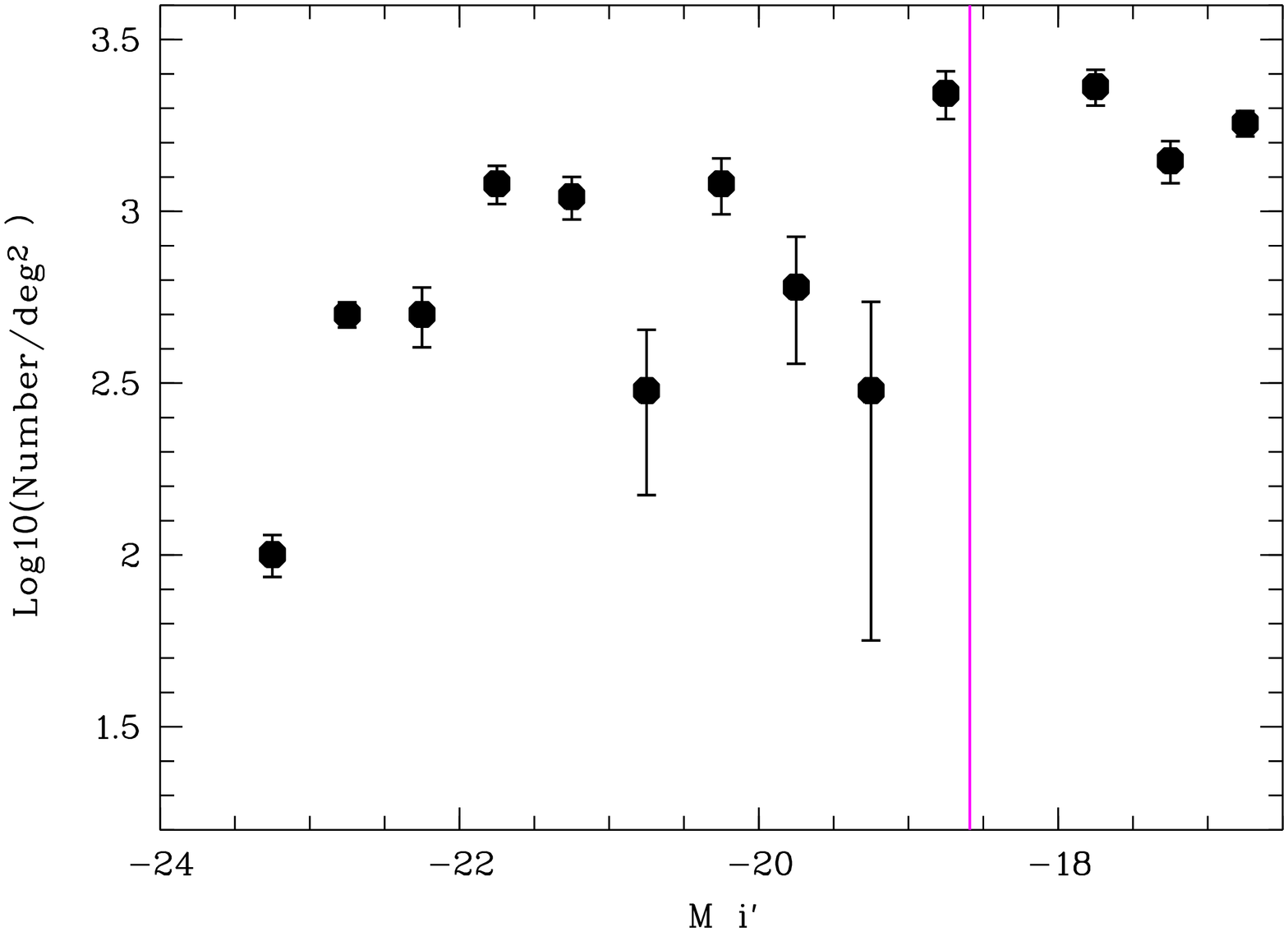}
    \caption{Luminosity functions for the whole galaxy population of
      D3-6 (top) and D3-43-S3 (bottom) per bin of 0.5 mag normalised
      to 1~deg$^2$. The vertical lines show the limits in absolute
      magnitude computed from the $i^\prime$=24 90\% completeness
      limit, applying the distance modulus and the mean K-correction
      in the D3-6 and D3-43-S3 fields. }
  \label{fig:fdlall}
  \end{center}
\end{figure}

We then compared the rest-frame absolute $i^\prime$ magnitude band
galaxy luminosity functions (in a 1.5 Mpc radius) of the two clusters
(Fig.~\ref{fig:fdlall}). We note that K-corrections are already taken
into account in the absolute magnitudes and this allows us to compare the
luminosity functions of the two main structures. To compare
the two luminosity functions, first we had to increase (0.07
towards the high values along the y-axis) the D3-6 values by 18$\%$ to take 
the fact that the two structures are at different redshifts into
account. We limited the luminosity functions to absolute magnitudes
brighter than $-18.5$ (see Fig.~\ref{fig:fdlall}). We then performed
10,000 realisations of the two luminosity functions within the error
bars and performed a Kolmogorov-Smirnov test each time. The
probability for the two luminosity functions to originate from the same
parent distribution was then found to be higher than 99$\%$ for 47$\%$
of the realisations, and higher than 75$\%$ for 97$\%$ of the
realisations. This clearly shows that the D3-6 and D3-43 luminosity
functions are not significantly different.

\begin{figure}[!ht]
  \begin{center}
    \includegraphics[angle=270,width=3.5in]{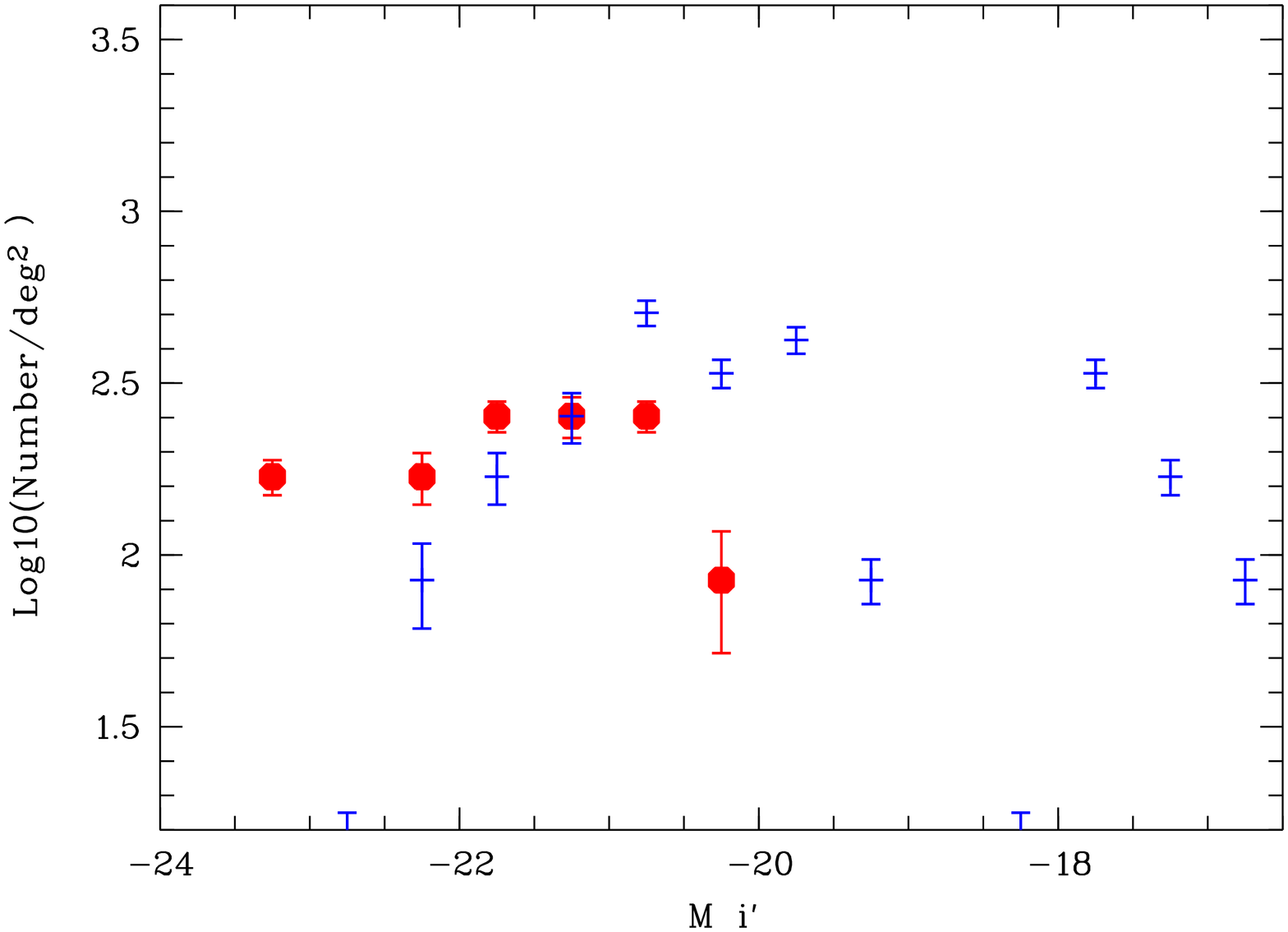}
    \includegraphics[angle=270,width=3.5in]{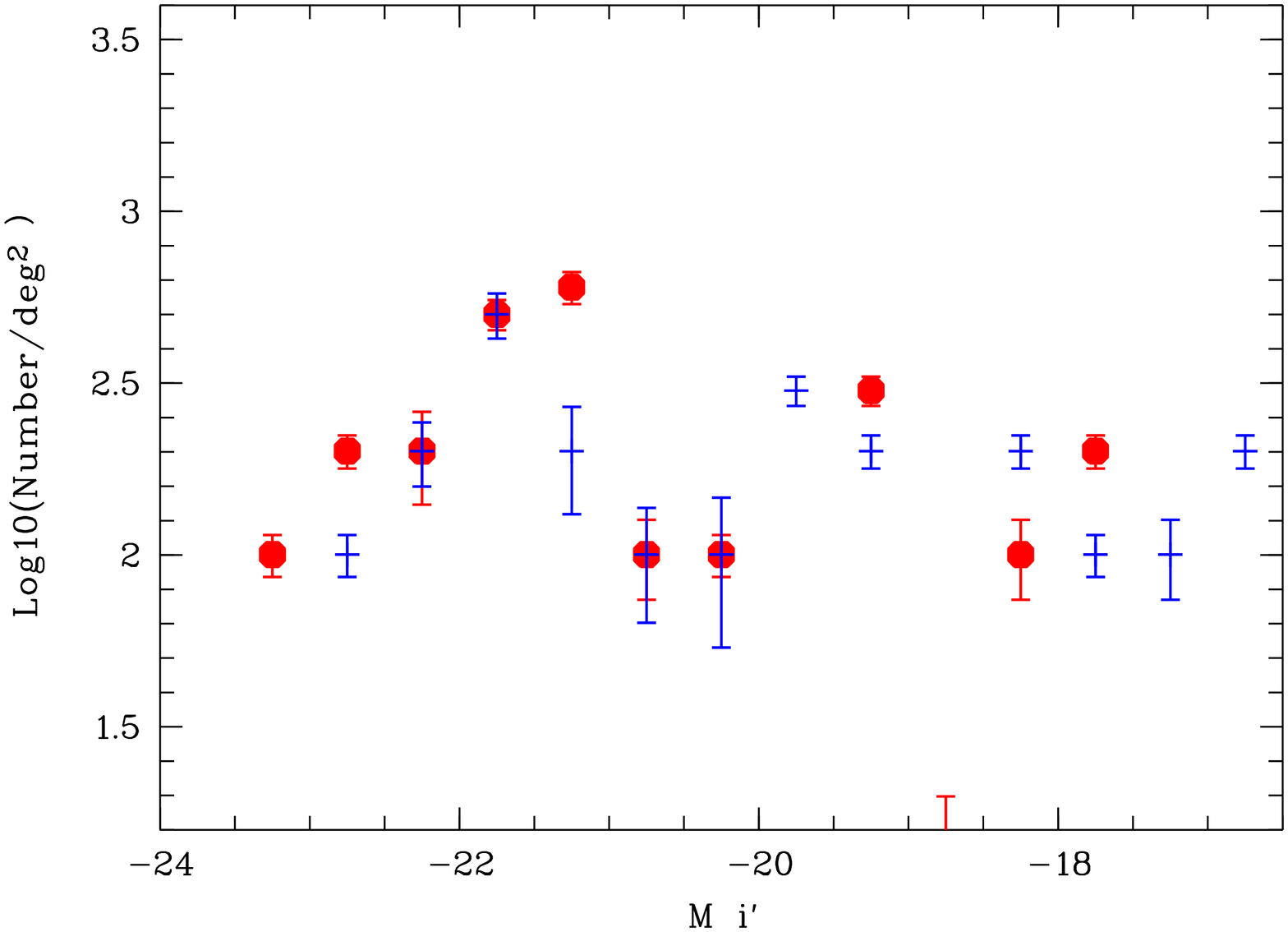}
    \caption{Luminosity functions for the galaxies with morphological
      types classified as later (blue crosses) and earlier (red disks)
      than Sb. Upper figure: D3-6; lower figure: D3-43-S3.}
  \label{fig:fdltype}
  \end{center}
\end{figure}

The next step is to consider the luminosity functions per
morphological type. This can be done by considering the modelled types
available in the photometric redshift catalogues. These types go from
1 (elliptical galaxies) to 31 (starburst galaxies). We divided our
galaxy sample into types earlier and later than Sb to
generate Fig.~\ref{fig:fdltype}. 

We performed the same Kolmogorov-Smirnov tests and
found that for the D3-6 cluster, the probability for the early and
late-type galaxy luminosity functions to originate from the same
parent distribution was higher than 99$\%$ for 0$\%$ of the
realisations, and higher than 75$\%$ for only 63$\%$ of the
realisations. In this case, we can conclude that the early and late-type 
galaxy luminosity functions in D3-6 are probably different, and
that this cluster clearly has more late-type galaxies than early-type
galaxies in the faint magnitude regime (this is quite obvious in
Fig.~\ref{fig:fdltype}).

  The same exercise yields different results for the D3-43-S3
  cluster. The probability that the early and late-type galaxy
  luminosity functions originate from the same parent distribution is
  higher than 99$\%$ for 42$\%$ of the realisations, and higher than
  75$\%$ for 100$\%$ of the realisations. In this case, we have
  similar luminosity functions for early and late-type galaxies (as
  seen in Fig.~\ref{fig:fdltype}).

\subsection{Star formation history in D3-6 and D3-43-S3}

\begin{figure}[!ht]
  \begin{center}
    \includegraphics[angle=270,width=3.5in]{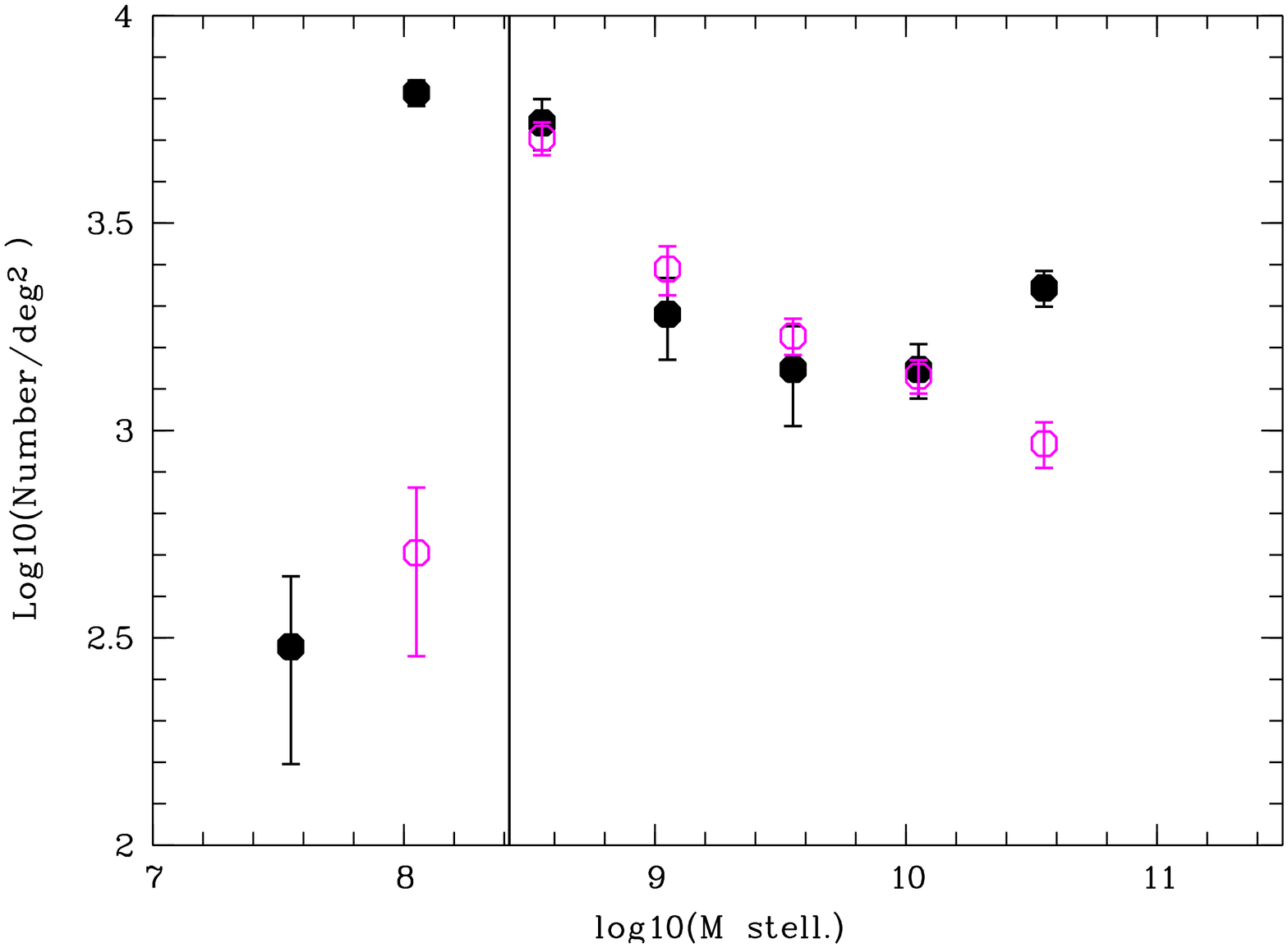}
    \includegraphics[angle=270,width=3.5in]{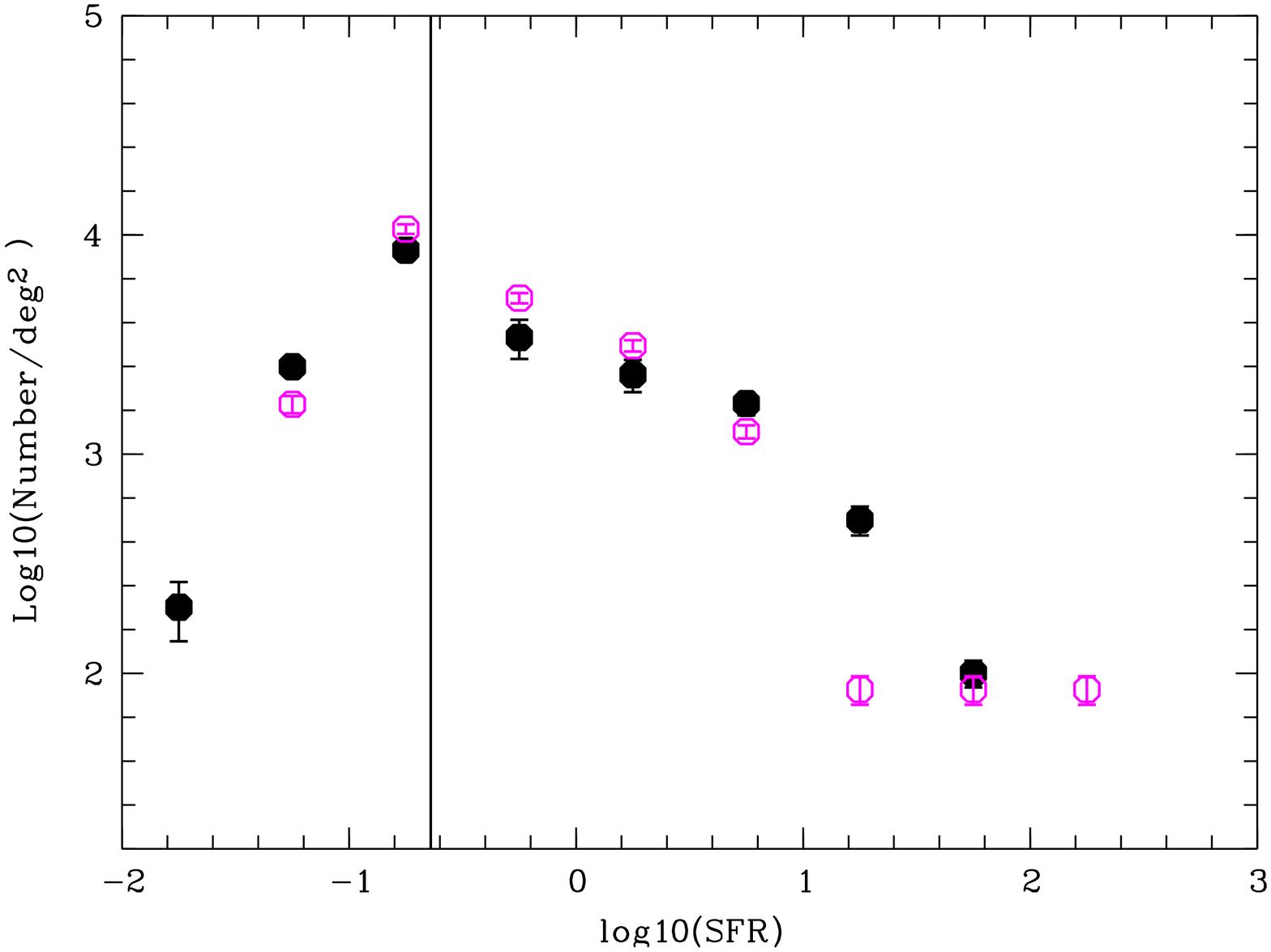}
    \caption{Upper figure: stellar mass functions for the galaxy
      populations of D3-6 (pink circles) and D3-43-S3 (black disks).
      Lower figure: SFR functions for the galaxy populations of D3-6
      (pink circles) and D3-43-S3 (black disks), in units of
      M$_\odot$/yr.  In both plots, the vertical lines show the values
      corresponding to the most stringent $i^\prime$ band absolute
      magnitude 90\% completeness level. }
  \label{fig:stellmass}
  \end{center}
\end{figure}

To investigate the stellar formation
  history in the galaxies of the considered clusters more precisely, we first
  computed galaxy stellar mass functions in D3-6 and D3-43-S3 (after
  statistical subtraction of the empty field values).  These are
  displayed in Fig.~\ref{fig:stellmass}. The two clusters exhibit very
  similar stellar mass functions. Even if D3-43-S3 is richer than D3-6
  for the highest stellar mass bin, a Kolmogorov-Smirnov test shows
  that the two global distributions (in the domain where the data are
  90\% complete) are the same. 

We also considered the star formation rates (SFR hereafter) of the
galaxies in D3-6 and D3-43-S3 (see Fig.~\ref{fig:stellmass}). A
  Kolmogorov-Smirnov test shows that the SFRs are very similar for
  both structures: probability greater than 99$\%$ for 49$\%$ of the
  realisations, and higher than 75$\%$ for 100$\%$ of the
  realisations.  Considering each bin of Fig.~\ref{fig:stellmass}
  separately, the only significant differences occur towards high
  values of the SFR, with D3-43-S3 having more galaxies with SFR of
  $\sim$16~M$_\odot$/yr (these galaxies could have  their SFR boosted
  by the ongoing merger taking place in D3-43-S3), and D3-6 having
  more galaxies with intense SFR ($\sim$150~M$_\odot$/yr).

  We estimated the SFR and stellar mass completeness levels in the
  following way: we selected the galaxies with a magnitude differing
  by less than 0.1 from the $i^\prime$ magnitudes at the 90\%
  completeness limit. We then computed the mean SFRs and stellar
  masses for these galaxies and considered that these values
  represented the SFR and stellar mass completeness levels.

\section{Conclusions}

With data obtained with GEMINI/GMOS and data taken from the literature
and NED, we spectroscopically confirmed two structures initially
detected with the AMACFI photometric redshift based cluster finder in
the CFHTLS D3 field.

The first structure, D3-43-S3, can be decomposed into two
substructures, each with a velocity dispersion of $\sim$350 and
$\sim$600~km~s$^{-1}$, which could be in the process of collapsing into a
large structure more or less along the line of sight. Since no X-ray
emission is detected, the interaction is probably not strong yet,
explaining the fact that we (barely) detect it in weak lensing (where we are
probably seeing the addition of the two structure masses along the line of
sight) but not in X-rays. 

The D3-6 structure is found to be a single structure at an average redshift
  $z=0.607$, with a velocity dispersion of 423~km~s$^{-1}$. It appears
  to be a relatively low-mass cluster.

We also show that D3-6 and D3-43-S3 have similar global galaxy
  luminosity functions, stellar mass functions, and SFR
  distributions. The only differences are that D3-6 exhibits a lack of
  faint early-type galaxies, a deficit of extremely high stellar mass
  galaxies compared to D3-43-S3, and an excess of very high SFR
  galaxies.

Besides the fact that this work adds two compact galaxy structures to
the strategical and still relatively poorly cluster-populated $0.5< z
< 1$ redshift range, it also shows the power of photometric redshift
based techniques to detect and study distant clusters (galaxy types,
SFR, etc.), provided that a large spectral coverage in the optical and
near infrared is available.  Combined-approach cluster surveys
(e.g. photometric redshifts to detect them and X-rays or weak lensing
to characterise them in terms of mass) are also crucial. EUCLID is the
perfect example of such a mission for clusters, since it will combine
weak lensing and cluster detections based on photometric redshifts.

\begin{acknowledgements}

  We would like to dedicate this paper to Alain Mazure, deceased in
  2013, and who was part of this work in its early stages. We
  gratefully acknowledge financial support from the Centre National
  d'Etudes Spatiales during many years. This work was also supported
  by the Brazilian agencies FAPESP and CNPq and benefited from the
  CAPES-COFECUB agreement number 711/11. We thank the referee for
  interesting comments.

  Based on observations obtained at the
    Gemini Observatory, which is operated by the Association of
    Universities for Research in Astronomy, Inc., under a cooperative
    agreement with the NSF on behalf of the Gemini partnership: the
    National Science Foundation (United States), the Science and
    Technology Facilities Council (United Kingdom), the National
    Research Council (Canada), CONICYT (Chile), the Australian
    Research Council (Australia), Minist\'erio da Ci\^encia,
    Tecnologia e Inova\c c\~ao (Brazil) and Ministerio de Ciencia,
    Tecnolog\'ia e Inovaci\'on Productiva (Argentina). This research
    has made use of the VizieR catalogue access tool at CDS,
    Strasbourg, France. This research has also made use of the
    NASA/IPAC Extragalactic Database (NED), which is operated by the
    Jet Propulsion Laboratory, California Institute of Technology,
    under contract with the National Aeronautics and Space
    Administration. Finally, this paper is based on observations
    obtained with MegaPrime/MegaCam, a joint project of CFHT and
    CEA/IRFU, at the Canada-France-Hawaii Telescope (CFHT), which is
    operated by the National Research Council (NRC) of Canada, the
    Institut National des Sciences de l'Univers of the Centre National
    de la Recherche Scientifique (CNRS) of France, and the University
    of Hawaii. This work is based in part on data products produced at
    Terapix available at the Canadian Astronomy Data Centre as part of
    the Canada-France-Hawaii Telescope Legacy Survey, a collaborative
    project of NRC and CNRS. 

\end{acknowledgements}

\clearpage

\appendix

\section{DEIMOS/GMOS redshift comparison}

\begin{figure}[!ht]
  \begin{center}
    \includegraphics[angle=270,width=3.5in]{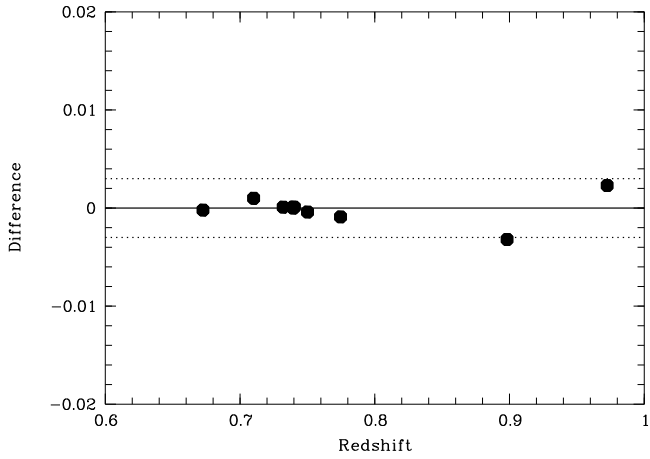}
    \caption{Comparison of Deep2 Keck/DEIMOS and GMOS/Gemini
      redshifts. The solid horizontal line symbolises the perfect
      agreement. The two dashed lines represent the 3$\sigma$ maximum
      uncertainty between the two redshift measurements taking the resolution of 
      the GMOS grism and the smoothing of the spectra into account.}
  \label{fig:zz}
  \end{center}
\end{figure}

We show in  Fig.~\ref{fig:zz} the difference between the GEMINI/GMOS redshifts 
and the KECK/DEIMOS DEEP2 redshifts as a function of redshift.

\section{Examples of spectra}

\begin{figure}[!ht]
  \begin{center}
    \includegraphics[angle=270,width=3.5in]{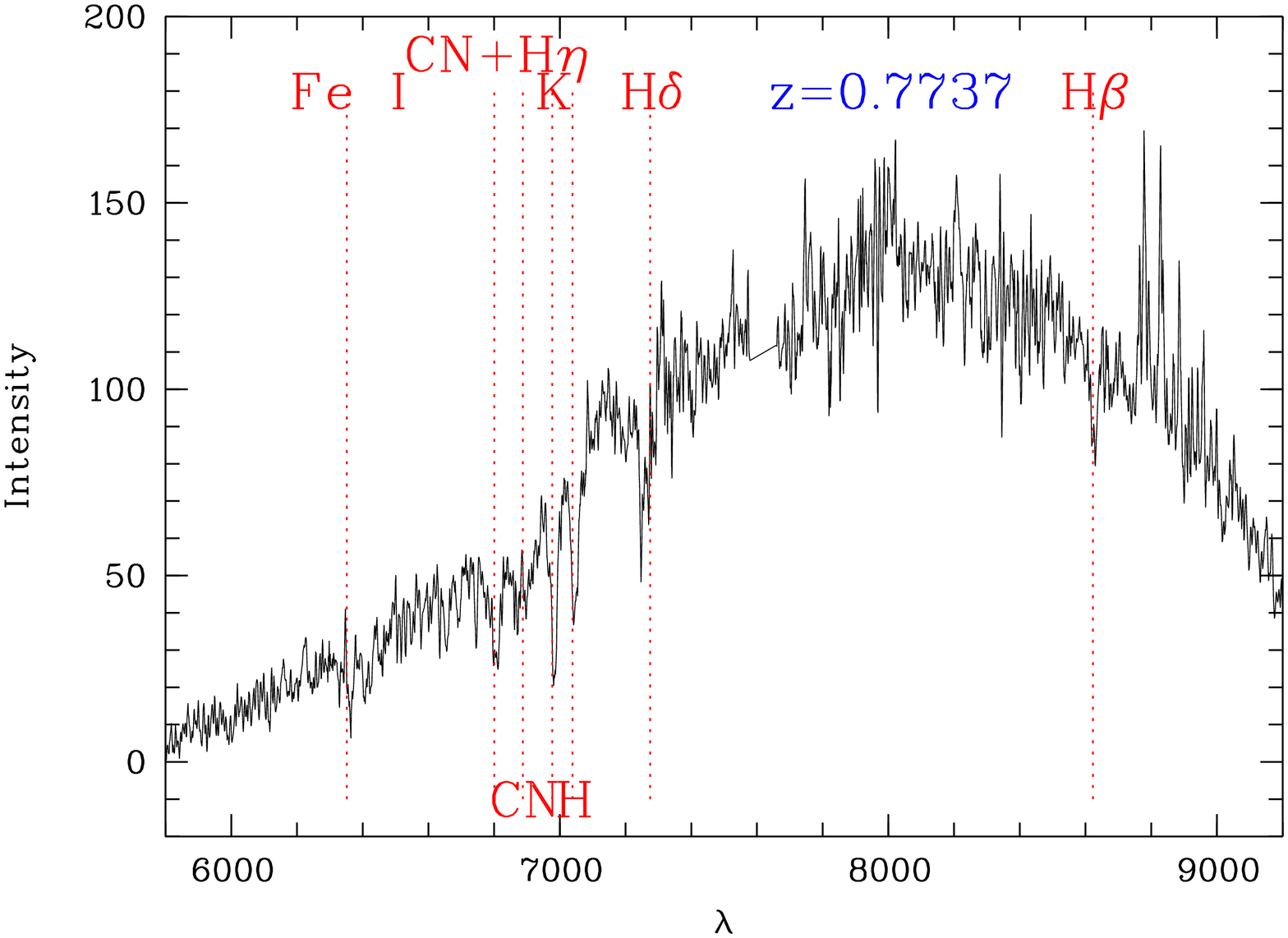}
    \includegraphics[angle=270,width=3.5in]{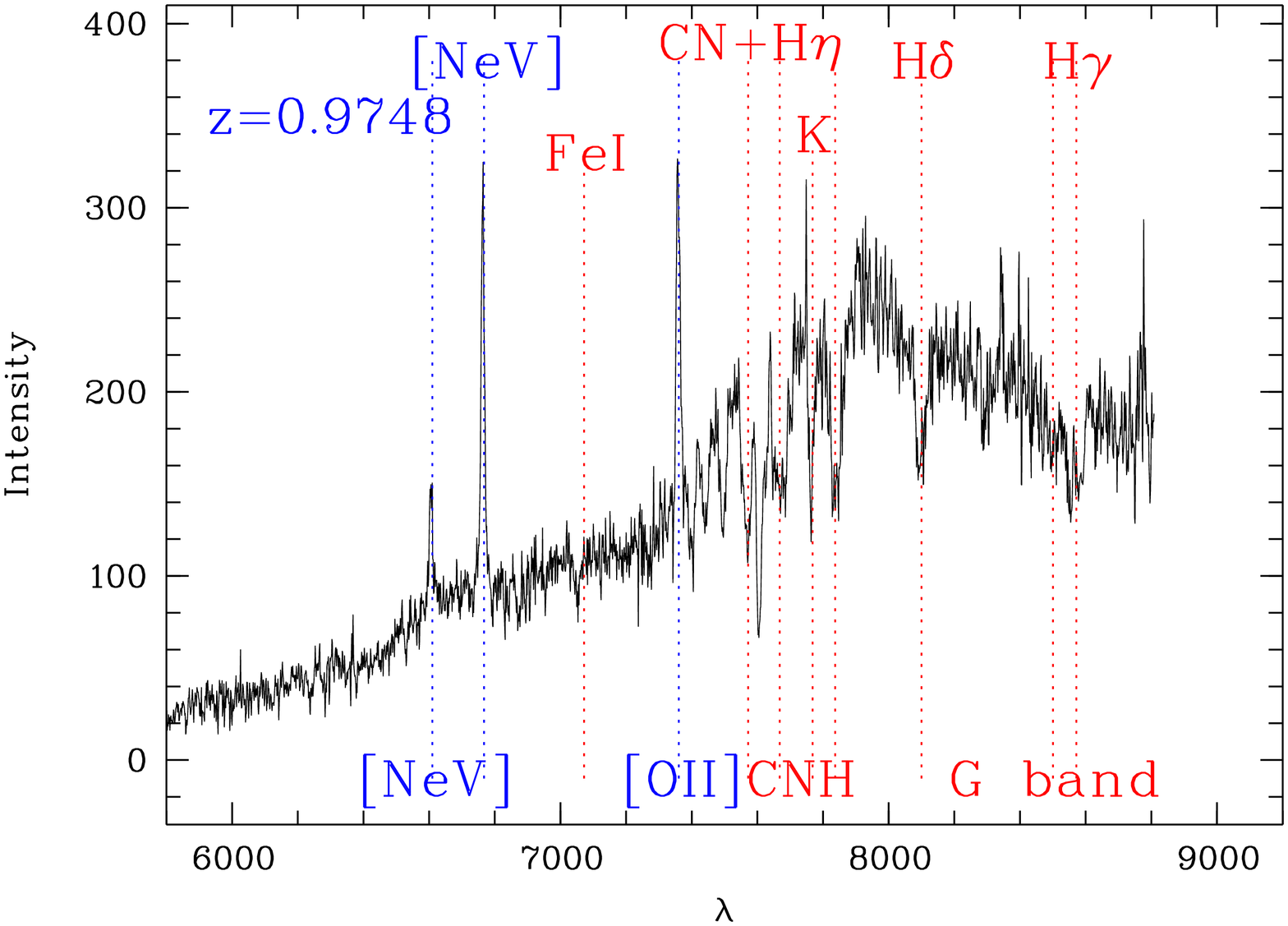}
    \caption{Two examples of Gemini/GMOS spectra with flags 4.  
Red labelled lines show absorption lines and blue labelled lines show emission lines.}
  \label{fig:example1}
  \end{center}
\end{figure}

\begin{figure}[!ht]
  \begin{center}
    \includegraphics[angle=270,width=3.5in]{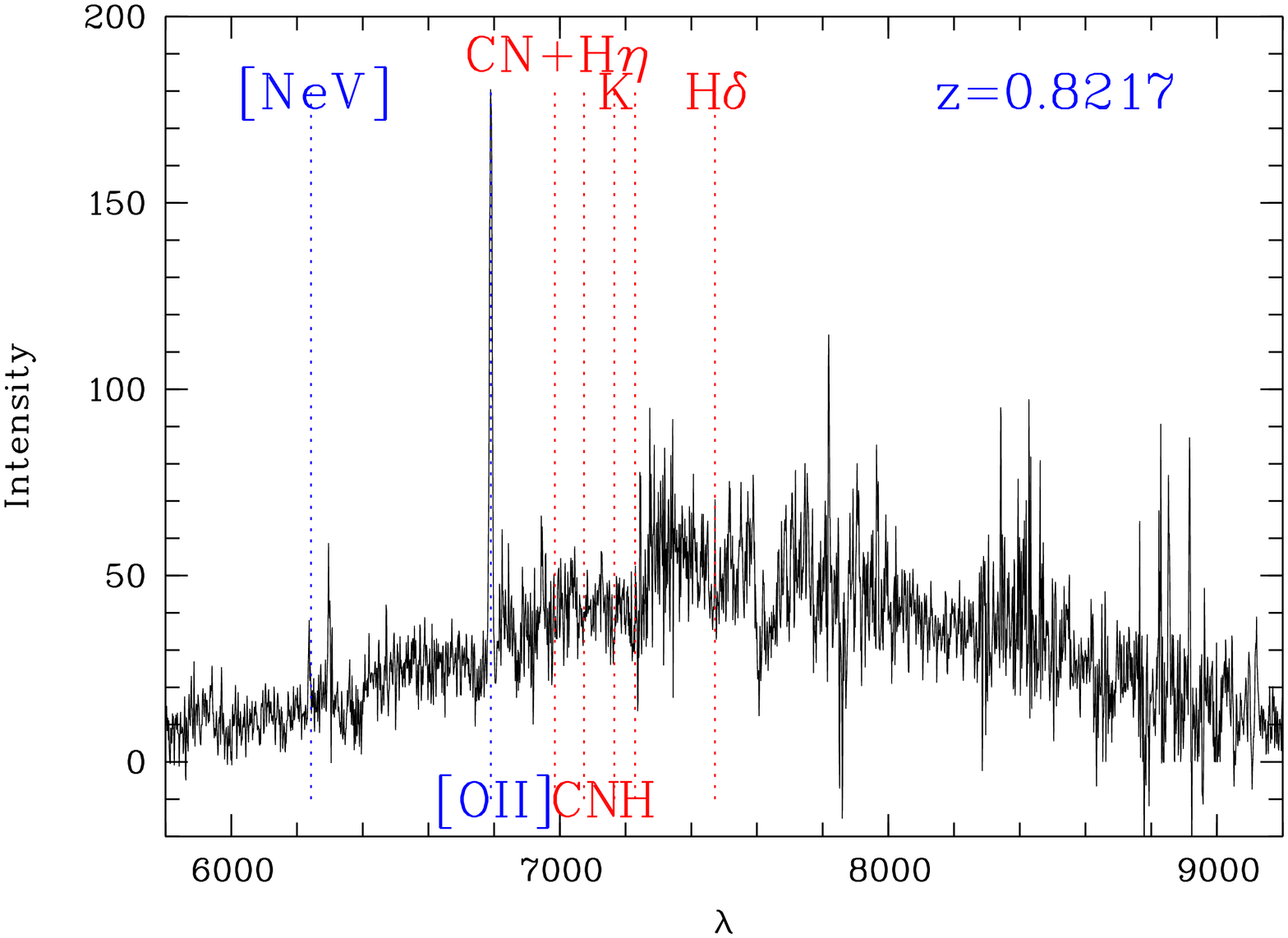}
    \includegraphics[angle=270,width=3.5in]{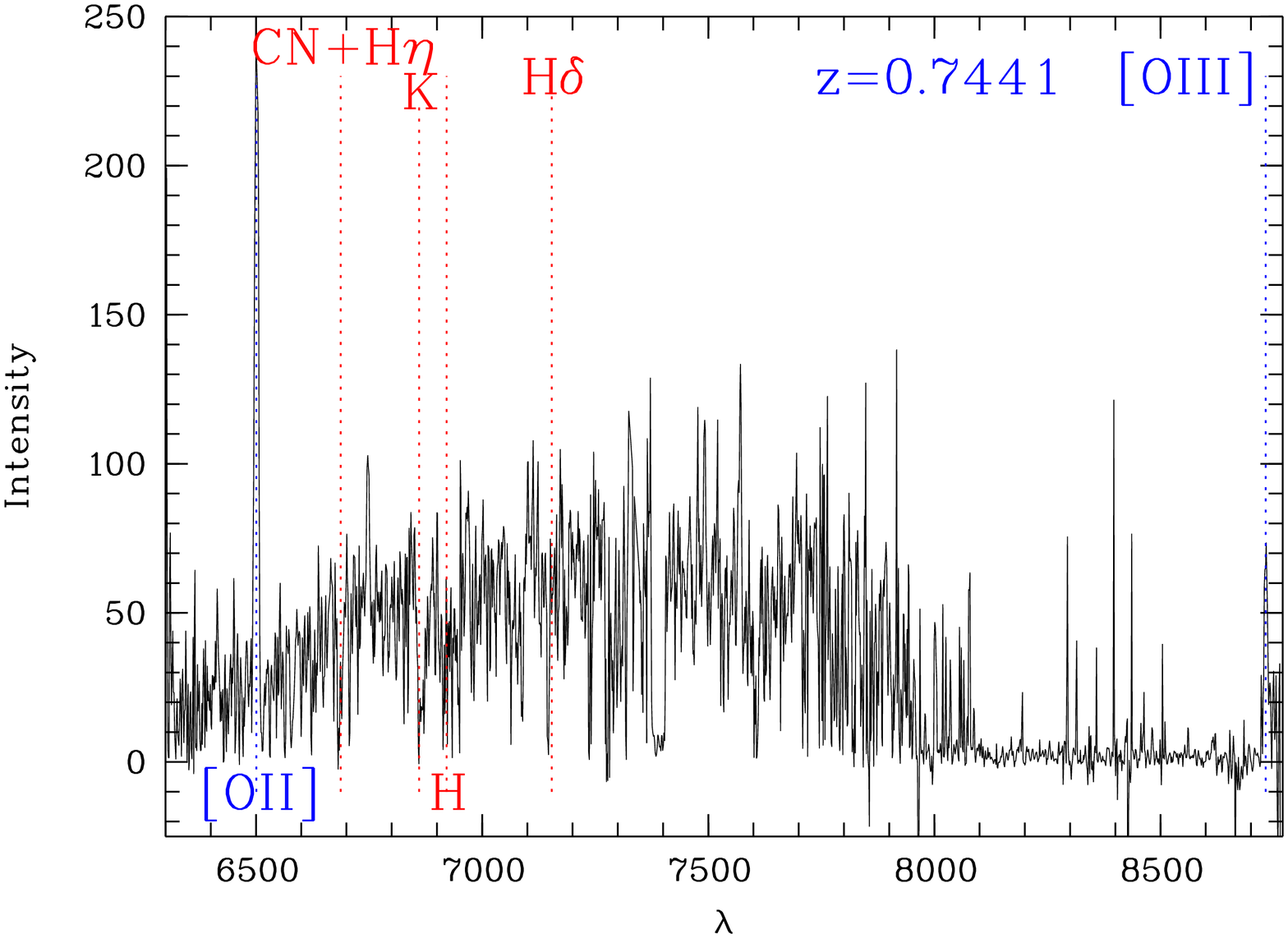}
    \caption{Two examples of Gemini/GMOS spectra with flags 3 and 2.  
Red labelled lines show absorption lines and blue labelled lines show emission lines.
From top to bottom: one spectrum with flag 3 and one spectrum with flag 2.}
  \label{fig:example2}
  \end{center}
\end{figure}

Below are four examples of spectra corresponding to flags 4, 3, and 2 in
  Figs.~\ref{fig:example1} and .~\ref{fig:example2}.

\end{document}